\documentclass[cmp]{svjour}

\usepackage{amsmath}
\usepackage{amsfonts,amssymb}
\usepackage{makeidx}
\usepackage{color,graphicx}	 

 \spnewtheorem*{rem}{Remark}{\it}{\rm}
 \spnewtheorem*{rems}{Remarks}{\it}{\rm}

\newcommand{\ZZ}{\mathbb{Z}}
\newcommand{\CC}{\mathbb{C}}
\newcommand{\NN}{\mathbb{N}}
\newcommand{\RR}{\mathbb{R}}

\newcommand{\tp}{\thinspace}
\newcommand{\la}{\langle}
\newcommand{\ra}{\rangle}
\newcommand{\supp}{{\rm supp}}
\newcommand{\sgn}{{\rm sgn}}

\newcommand{\erfc}{\mathrm{erfc}}

\newcommand{\idty}{\mathbf{1}}

\newcommand{\vx}{\mathbf{e}_x}
\newcommand{\vy}{\mathbf{e}_y}

\renewcommand{\Re}{\mathrm{Re}\thinspace}	
\renewcommand{\Im}{\mathrm{Im}\thinspace}	
\newcommand{\intd}{\mathrm{d}}			

\newcommand {\curl}{{\rm curl}}

\newcommand{\ii }{\mathrm{i}}		
\newcommand{\vect}[1]{\boldsymbol{#1}}	
\newcommand{\ex}{\vect{e}_\mathrm{x}}	 

\newcommand{\taothou}{\vect{m}^{\mathrm{TT}}}	
\newcommand{\half}{\frac{1}{2}}  
\newcommand{\perm}{\mathcal{S}}		

\newcommand{\ml}{\ell}  

\title{Symmetry breaking in Laughlin's state on a cylinder}

\author{ S. Jansen\inst{1}
	 \and  E. H. Lieb\inst{2}
	 \and R. Seiler\inst{1}}
\institute{Institut f\"ur Mathematik, Technische Universit\"at Berlin, Str. des 17. Juni 136, 10623 Berlin, 
	Germany
	\and Department of Mathematics, Princeton University, P. O. Box 708, Princeton, NJ 08544, USA
	}

\mail{Sabine Jansen, \email{jansen@math.tu-berlin.de}\\
\copyright 2008 by the authors. This paper may be reproduced, in its entirety, for non-commercial purposes.}

\begin{document}

\maketitle
\begin{abstract} 
	We investigate Laughlin's fractional quantum Hall effect wave function 
	on a cylinder. We show that it displays translational symmetry breaking 
	in the axial direction for sufficiently thin cylinders.
 	At filling factor $1/p$, the period is $p$ times 
	the period of the filled lowest Landau level. The proof uses a connection with 
	one-dimensional polymer systems and discrete renewal equations. 
\end{abstract}

\tableofcontents

\section{Introduction}

Our goal in this paper is to investigate the properties of a quantum
mechanical wave function of very many variables known as the Laughlin
function \cite{l83}. It arises as an explanation of a curious and unexpected
phenomenon known as the fractional quantum Hall effect (FQHE) \cite{tsui83}.

It is not our intention to discuss the physics of FQHE but
rather to discuss the mathematical problems that arise when the
Laughlin function is appropriately modified for a two-dimensional
cylindrical geometry \cite{thousurfsci}. The interesting  point here is that although
the setup is seemingly translation invariant (in the direction of the
cylinder axis) the resulting one-particle density $\rho$ is not
invariant but has a non-trivial periodicity of length $p$ (in natural
units), where $p$ is an integer related to the 
so-called ``filling fraction''.  
The main content of this paper is the rigorous verification
of this periodicity when the cylinder radius is small compared to the
natural length. The periodicity has a natural interpretation 
in the context of classical Coulomb systems and is manifest 
in the periodicity of a closely related ``quantum polymer''~\cite{jls}. 
The conjecture of periodicity for all radii remains open.\\

Let us just say a few words about the function and why it was 
proposed.  Imagine a hollow circular cylinder in $\RR^3$ of radius $R$
on whose surface $N$ electrons reside. Each electron has a coordinate
$(x,y)$, with $x\in \RR$ denoting the distance parallel to the
cylinder axis and $y \in [0,2\pi R)$ denoting the angular coordinate.
We will often use $z=x+\ii y$ to denote the coordinate instead of
$(x,y)$. 
There is a magnetic field of magnitude $\vect{B}= \curl \vect{A}$
 perpendicular to the
surface.  Since magnetic fields must be divergence free we can think
of this situation as one in which there is a (unobserved) flux through
the hollow cylinder that leaks out through the surface with a constant
flux through the cylindrical surface on which the electrons reside (see also~\cite{west}, Sect.~III).
 If we pay attention only to the surface the physical situation is
translation invariant but if we pay attention to the flux in the
hollow core we can measure our $x$ coordinate by measuring the total
flux in the circular cross section at $x$. The magnetic vector
potential, $\vect{A}$, evaluated on the surface, is given by Stokes' theorem
(assuming that $\vect{A}$ has no component in the axial direction) as
$\vect{A}(x,y) = (0, Bx +\phi/(2\pi R))$, which is obviously periodic in the $y$
 direction. Our vector potential $\vect{A}(z)$ is determined
only up to a gauge parametrized by $\phi$. 

With this choice the one particle Schr\"odinger operator is 
\begin{equation} \label{laugh}
  H  =\frac{1}{2}\bigl(\vect{p}-\vect{A}(z)\bigr)^2
	= \frac{1}{2}((-i\partial_x)^2 + (-i\partial_y - Bx -
 \frac{\phi}{2\pi  R})^2).
\end{equation}
In our  units  Planck's constant $\hbar$, the mass and the
charge of the electron are $1$.  To simplify notation $B$ actually
 denotes
the magnetic field divided by the speed of light, and in the following we will choose 
units in such a way that $B=1$. Our operator $H$ should be seen as the part of a  
three-dimensional Hamiltonian which acts on the axial 
and the angular parts of the wave function.\footnote{In cylindrical coordinates, with 
$r,\theta,x$ the radial, angular and axial coordinates, the part of the Laplacian that we are 
interested in is $r^{-2}\partial_\theta^2 + \partial_x^2$. Fixing $r=R$ and choosing 
$y=R \theta$, this becomes $\partial_y^2+ \partial_x^2$: the differential symbol is $R$-independent.}
 The vector potential $\vect{A}$ is the restriction to the cylinder surface of some 
three-dimensional vector potential that gives rise to a magnetic field as sketched 
in~\cite{west}, Fig.~3. 


In the chosen units, the ``magnetic length'' $\ml =(\hbar /eB)^{1/2}$ 
equals $1$. An important role will be played by the dimensionless parameter 
$$
	\gamma = \ml  / R = 1/R.
$$

Here we come to the crucial point that the usual gauge invariance
(meaning that we can replace $\vect{A}(z)$ by $\vect{A}(z) +\nabla \tilde\phi(z)$
 and a
wave function
$\Psi(z) $ by $\Psi (z)\exp[\ii \tilde\phi(z)]$) is not generally allowed.
Gauge changes $\exp(\ii \tilde \phi(z))$ have to respect the $2\pi R$-periodicity 
in the $y$ direction. 
Consequently, the number $\phi$ in $H$ cannot be changed by
a gauge transformation unless the change is a multiple of $2\pi$.

This lack of complete gauge symmetry is intimately related to the
breaking of translational symmetry.  If we translate in the axial
direction by $\delta$ then $x\to x+\delta$, the Hamiltonian~(\ref{laugh})
with this replacement is generally {\it not} gauge equivalent to the
original one. That is, changing $\vect{A}$ by $\nabla \tilde\phi$ with
$\tilde\phi = \delta y$ is not allowed -- except for the special cases
$\delta =n/R= n\gamma$ with $n \in \ZZ$. Hence the system possesses a
discrete translational symmetry.

The 
discrete translational invariance and the effect of changing $\phi$ can be nicely 
read off the eigenfunctions of $H$. The eigenfunctions belonging to the lowest eigenvalue 
 are all of the form 
\begin{equation*}
	\psi(z) = f(z) \exp\bigl[ -(x+\frac{\phi \gamma}{2\pi})^2/2\bigr]
\end{equation*}
where $f$ is an entire function. To respect $2\pi R$ periodicity, the eigenfunction 
must be a superposition of functions
\begin{equation}
 	\exp(\gamma n z)\exp \bigl[ -\bigl(x+\frac{\phi \gamma}{2\pi}\bigr)^2/2\bigr]
	\propto \exp(\ii n\gamma y) \exp \bigl[ -\bigl(x - (n-\frac{\phi}{2\pi})\gamma\bigr)^2/2
	\bigr], \label{lowl}
\end{equation}
with $n\in \ZZ$. These functions are essentially Gaussians of $x$; their 
centers form a lattice with spacing $\gamma$, and changing $\phi$ amounts to a uniform 
shift of the centers. From now on, we set the parameter $\phi$ to $0$. 

 If we  now consider a finite
cylinder of length $L$ 
 then we impose the `physical' boundary
condition that we restrict functions to the interval $0\leq x\leq L$ and consider 
only ground state eigenfunctions with centers inside that 
interval.\footnote{These are the ground state eigenfunctions of the Hamiltonian~(\ref{laugh})
with chiral boundary conditions \cite{aans}. }
Thus, for a finite cylinder, the dimension of this lowest energy space - known as the lowest 
Landau band - equals the 
``number of fluxes'' $RL$, i.e., the total flux $2\pi R L B$ divided by the flux quantum $2\pi$ 
(remember $B=1$).

If there are some number $N$ of electrons 
that repel each other via a Coulomb interaction we 
have to construct a wave
 function
of $N$ variables $z_1,z_2, ...,z_N $ that  minimizes the total
  energy. 
The kinetic part is minimal if we choose sums of products of
 one-particle
functions from the lowest  Landau band. In other words our wave
 function must
have the form 
\begin{equation*}
   P\bigl(\exp(\gamma z_1),  \exp(\gamma z_2), \dots ,
	\exp(\gamma z_N)\bigr) \exp(-\sum_{j=1}^N x_j^2/2)
\end{equation*}
where $P$ is a
polynomial in $N$ variables of degree at most $RL$ in each variable
 separately. 
Furthermore, the Pauli exclusion principle for fermions demands that
 this
polynomial be  antisymmetric, i.e., that it change sign if any two
 variables
are interchanged. If the particles are bosons, instead, then $P$ must
 be
symmetric. 

We are interested in the cases that $N=RL/p$ with $p=1,2,3,...$, in
 which case
we say that the lowest Landau band is filled to a fraction $1/p$.  The function 
\begin{equation} \label{laufunct:eq}
	\Psi_N(z_1,...,z_N)= \kappa_N \prod_{1\leq j<k\leq N} \bigl(\exp(\gamma z_k)-\exp(\gamma 
		z_j)\bigr)^p \exp(-\sum_{j=1}^N x_j^2/2)
\end{equation}
satisfies all the stated conditions for
fermions (when $p$ is odd) and for bosons (when $p$ is even). 
Later, it will be useful to fix the multiplicative factor $\kappa_N$ as in 
 Eq.~(\ref{multconst:eq}) below. The factor $\kappa_N$ does not 
affect the one-particle density. 
Our choice of $\kappa_N$ makes $\Psi_N$ 
the power of a Slater determinant of functions that resemble the lowest Landau level basis functions 
(see Eq.~(\ref{powdet:eq})), and considerably 
simplifies subsequent statements on normalization constants.

This function was invented by Laughlin \cite{l83} for the disk geometry
and later adapted to the cylinder geometry \cite{thousurfsci}, and is considered 
to be a good approximation to the true ground state. Its distinctive feature is that 
it is small when two particles $z_j$ and $z_k$ are close together, thereby 
making the interaction energy small. \\

In this article, we analyze the periodicity in the axial direction 
of the state $|\Psi_N\ra \la\Psi_N |$ 
 in the limit $N\rightarrow\infty$ 
for fixed radius $R$ and filling factor $1/p$. 
Particular emphasis will be put on the 
one particle density
\begin{equation*}
	\rho_N(z)= \frac{N}{C_N}
		\int_{\RR^{N-1}}\int_{[0,2\pi \gamma^{-1}]^{N-1}} 
			\bigl|\Psi_N (z,z_2,..,z_N)\bigr|^2 \intd y_2 .. \intd y_N \intd x_2.. \intd x_N, 
\end{equation*}
where $C_N = ||\Psi_N||^2$ is the $L^2$-norm squared with respect to integration over 
$(\RR\times [0,2\pi \gamma^{-1}])^N$. Notice that we integrate over infinite cylinders 
but keep referring to $\Psi_N$ as a finite cylinder state. The reason is that the wave function 
vanishes exponentially fast outside the finite cylinder. For $p=1$, the wave function 
is a simple Slater determinant of basis functions~(\ref{lowl}), the density is a sum of 
equally weighted Gaussians $\exp[-(x-n\gamma)^2]$ (see Eq.~(\ref{density:eq})), 
and changing the domain of integration to a finite cylinder does not affect the density in the middle of the 
cylinder. 

 The origin  of the analysis of the density's periodicity 
is an observation for the filled lowest Landau level: for $p=1$, the density 
is periodic with minimal period $\gamma$. This contrasts with the constant 
density of the filled Landau level in the symmetric gauge and reflects the 
discrete translational symmetry of the Hamiltonian~(\ref{laugh}). 

When $p>1$,  
Laughlin's state does not possess the full discrete 
symmetry of the Hamiltonian~(\ref{laugh}) any longer, hence the symmetry is broken.
However, the state is still periodic, 
with minimal period $p\gamma$, i.e., $p$ times the period of the filled Landau level. 
This is rigorously shown here for cylinders with sufficiently small radius (large $\gamma$);
numerical results \cite{swk04} for $p=2$ and $p=3$ suggest that the result actually holds 
for all values of the cylinder radius. Related, finite~$N$ results have been obtained in \cite{rh94}.
Related considerations for Laughlin type functions on tori, introduced in \cite{hr85}, 
can be found in \cite{leeleinaastorus,bk}. In \cite{leeleinaastorus},
the amplitude of oscillations of the density for torus Laughlin-type functions is 
investigated numerically, as a function of the cylinder radius. 

It follows from our results that the density and hence, the amplitude of $3\gamma$-oscillations,
 are analytic function of $\gamma$ when $\gamma$ 
is sufficiently large. This implies that the only possibility for the oscillation amplitude to vanish on 
some small $\gamma$ interval is that a phase transition occurs as $\gamma$ is decreased, 
see \cite{k74} for a similar argument. 
(With ``phase transition'' we mean that either there is no longer a well-defined unique thermodynamic limit 
or there is a unique thermodynamic limit but the density ceases to be analytic at some point.) 
\cite{bk} argue that no such phase transition occurs, but a rigorous proof is still lacking.

Due  to the picture mentioned before, translation along 
the cylinder axis can be reinterpreted as a change of flux $\phi$. One flux unit ($2\pi$)
corresponds to a translation by one natural length unit ($\gamma$). Hence the many
electron system is invariant under a change of $p$ flux units. In the
picture of Laughlin's original article on the integer Hall effect \cite{l81} this
corresponds to a transport of one electric charge from one edge to the
other of the cylinder for every adiabatic change by $p$ flux units.
This is related to a heuristic argument given in \cite{tw84}, suggesting 
that the periodicity is needed in order to reconcile Laughlin's IQHE article with 
fractional quantum Hall conductances. 

Another interpretation 
of the periodicity is that the $N$ particles in a cylinder of
length $pN$ are not arranged randomly  but,  more like a
one-dimensional crystal, they are separated by a distance of $p$ units. 
This interpretation is of interest from the point of view of Laughlin's plasma analogy 
\cite{l83}: $|\Psi_N|^2$ is proportional to the Boltzmann weight of a classical 
jellium system placed on a cylinder (see also~\cite{jls}, Sect.~3).
The factor $\exp(-\sum_j x_j^2/2)$ in the wave function arises from 
the potential created by a neutralizing background while the polynomial in 
$\exp(\gamma z_j)$ accounts for the repulsive interaction between charged particles. 
One-dimensional jellium systems always display
periodicity, i.e.,  ``Wigner crystallization'', as was shown in \cite{k74,bl75}.
For jellium tubes, the symmetry breaking is well-known 
at coupling constant $\Gamma =2$, corresponding to the filled Landau level, see 
\cite{cfs83,jl01}. The periodicity has been conjectured to hold for other values of 
the coupling constant \cite{agl01}. Our results give a partial proof of this conjecture. \\

Our  result on the periodicity 
comes in several parts: a preliminary observation 
(Lemma~\ref{foutrans:lem}), holding 
for all values of $\gamma$, is that the one particle density can never be constant, and if it is 
periodic, the period must be a multiple of $\gamma$. Theorem~\ref{thermolim:theo}, valid for large $\gamma$,
states that 
the Laughlin state has a unique thermodynamic limit as $N\rightarrow \infty$, and the limiting 
state has $p\gamma$ as one of its periods. Theorem~\ref{symbreak:theo} says that $p\gamma$ is 
actually the smallest period. We also prove, for large $\gamma$, that the state is 
mixing with respect to (magnetic) translations along the direction of the cylinder axis 
(Theorem~\ref{cluster:theo}).

\begin{figure}
	\resizebox{8cm}{!}{\input{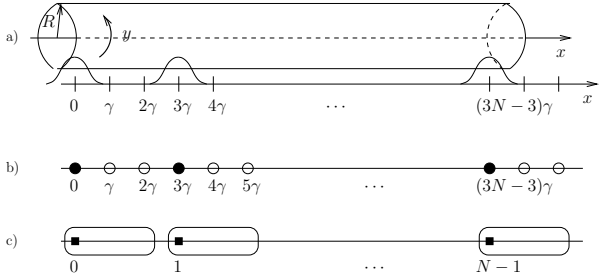}}
	\caption{\label{fig:intro} 
		a) The Tao-Thouless state for $p=3$ and $N$ particles is made up of lowest Landau band 
		functions~\ref{lowl}  with Gaussian centers $0$,$3\gamma$,..,$(3N-3)\gamma$.
		b) This is visualized as a succession of $N$ times the sequence ``filled circle - white 
		circle - white circle''.
 		c) These $N$ blocks are associated with a monomer partition of $\{0,...,N-1\}$.
		 }
\end{figure}

The crucial idea  for the proof of our results is the representation 
of Laughlin's wave function as a ``quantum polymer''~\cite{jls}. This means that $\Psi_N$ 
is written as a sum of functions, each associated with a partition of 
$\{0,..,N-1\}$. One of these functions is the Tao-Thouless function \cite{tt83}, which is the antisymmetrized 
tensor product of lowest Landau band functions centered at multiples of $p\gamma$. The Tao-Thouless 
function can be interpreted as a simple monomer, see Fig.~\ref{fig:intro}. 
The polymer expansion is explained 
in Subsects.~\ref{model:sec} and~\ref{poly:sec}. 
The associated polymer system has a built-in form of translational invariance 
that is at the origin of the $p\gamma$-periodicity in the limit $N\rightarrow \infty$. 

One consequence of the polymer representation is a recurrence relation for 
the normalization constant $C_N$, known as a (discrete) renewal equation. 
In Subsect.~\ref{norm:sec}, we show that the associated renewal 
process has finite mean for sufficiently large $\gamma$. We use this important 
technical result in Subsects.~\ref{thermo:sec} and~\ref{symclu:sec} to prove our 
results on Laughlin's function. 

At this point, let us mention that 
plasma analogy sheds some light not only on the density's periodicity, but also on the 
recurrence relation: indeed, in the context of jellium systems, a renewal equation 
already appeared in \cite{len61}, while a renewal inequality showed up in \cite{leli}.

In Sect.~\ref{solv:sec}, we investigate a solvable model that arises as a simplification 
of Laughlin's function. In this model, only monomer-dimer systems occur. 
This allows us to introduce in a simpler setting the ideas 
described above.

 
\section{A solvable model} \label{solv:sec}

As a preparation to 
the examination of Laughlin's wave function, 
we analyze a modified wave function. It is derived from Laughlin's wave function 
by replacing Gaussians with functions of compact support. 
If the support is small, the normalization constants and the
one-particle density can be computed explicitly (Prop.~\ref{modfunct:prop}),
and it is easy to check that the thermodynamic limit of the one-particle
density is a periodic function (Corollary~\ref{thermolim:cor}). 

The proof of these results relies on the representation of the modified
wave function as a sum over monomer-dimer partitions
(Lemma~\ref{solvpoly:lem}), leading to a recurrence relation of second order
for the normalization constants, see Eq.~(\ref{solvrecurr:eq}). 

The derivation of the modified wave function starts from the well known representation of 
Laughlin's wave function 
 as the $p^\mathrm{th}$ power of a Vandermonde 
determinant~\cite{d93,fgil94} times a Gaussian weight. The Gaussians can be absorbed into 
the determinant and we can complete the squares. For $N=2$ particles and 
$p=3$, this gives 
\begin{align*}
	&\bigl( \exp(\gamma z_2) - \exp(\gamma z_1) \bigr)^3 \exp\bigl(-\frac{1}{2}(x_1^2+ x_2^2) \bigr)\\
		&\quad = \det \begin{pmatrix}
		        \exp(-x_1^2/6) & \exp(-x_2^2/6) \\
			 \exp(\gamma z_1 - x_1^2/6) &\ \  \exp(\gamma z_2 - x_2^2/6) 
		       \end{pmatrix}^3\\
		&\quad = \exp(\frac{9}{2}\gamma^2) \det \begin{pmatrix}
		        \exp(-x_1^2/6) & \exp(-x_2^2/6) \\
			 \exp\bigl[\ii \gamma y_1 - (x_1-3\gamma)^2/6\bigr]
			 &\ \ \exp\bigl[\ii \gamma y_2- (x_2-3\gamma)^2/6\bigr]
		       \end{pmatrix}^3.
\end{align*}
More generally, with a suitable choice of $\kappa_N$ (see Eq.~(\ref{multconst:eq}) below),
 Laughlin's function becomes 
\begin{equation}\label{powdet:eq}
 	\Psi_N(z_1,..,z_N):=\frac{1}{\sqrt{N!}} 
			\Bigl( \det\bigl(\varphi_{k-1} (z_j)\bigr)_{1\leq k,j\leq N} \Bigr)^p
\end{equation}
where 
\begin{equation*}
	\varphi_k(z)
	= \frac{1}{(2\pi \gamma^{-1}\sqrt{\pi})^{1/2p}} \exp(\ii k\gamma y)
		\exp\bigl( -(x-pk\gamma )^2/2p\bigr).
\end{equation*}
Notice that, up to the multiplicative constant, the quantity $\gamma$ appears
 only in combination with the integer $k$. 
Hence it suffices to consider the case $\gamma =1$. The 
$\gamma$-dependence can be restored by replacing in the forthcoming formulas
integers $k\in \ZZ$ by $k\gamma$.
To simplify the formulas further, we will limit ourselves to $p=3$. 

Now we replace the Gaussian in formula~(\ref{powdet:eq}) by an even measurable function 
$f$ of compact support. Let
\begin{equation} \label{rootbasefunct-mod:eq}
	\phi_k(z):= \exp(\ii k y) f(x-3 k), \quad(k\in \ZZ).
\end{equation}
For $N\in \NN$, we define the modified wave function $\Phi_N$ by
\begin{equation} \label{modfunct:eq}
	\Phi_N(z_1,..,z_N):=\frac{1}{\sqrt{N!}} 
		\Bigl( \det\bigl(\phi_{k-1}(z_j)\bigr)_{1\leq k,j\leq N} \Bigr)^3.
\end{equation}
It turns out that the one-particle density can be computed exactly when only
neighbors overlap: $\phi_k \phi_m = 0$ if $|k-m|\geq
2$. This can be achieved by choosing a function $f$ with a compact support
contained in $[-3, 3]$, so that 
\begin{equation} \label{nearestneighbor:eq}
	f(\cdot) f(\cdot -  3n ) = 0 \ \text{for}\ |n|\geq 2.
\end{equation}
The main idea in this section 
is to use the nearest neighbor overlap condition to simplify the
expansion of the determinant power~(\ref{modfunct:eq}). 
Because of the third power, the expansion of Eq.~(\ref{modfunct:eq}) can be written 
as a sum of wedge products \footnote{We use the normalization 
 	$(f\wedge g)(z_1,z_2)= \frac{1}{\sqrt{2}} \bigl( f(z_1)g(z_2)- f(z_2) g(z_1)\bigr)$.}
of functions $\phi_k(z) \phi_m(z) \phi_n(z)$. The overlap 
condition restricts $k,m,n$ to $k=m=n$, $k=m$, $n=k+1$, etc.
We define an orthonormal set of functions $\psi_m$, labelled by their 
$y$ momentum $m\in \ZZ$, by:
\begin{alignat*}{2}
	\phi_k^3 (z)&= \sqrt{\alpha_1} \psi_{3k}(z),&\quad
	 \phi_k^2(z) \phi_{k+1}(z) &= \sqrt{\beta_2} \psi_{3k+1}(z)\\ 
	& & 	\phi_k(z) \phi_{k+1}^2(z) &= \sqrt{\beta_2} \psi_{3k+2}(z),
\end{alignat*}
The wave function $\psi_m(z)$ describes a particle
localized 
around $x=m$. The constants $\alpha_1$ and $\beta_2$ are suitable 
positive normalization constants. 

The normalization of $\Phi_N$ and the one-particle density are easily 
expressed in terms of $\alpha_1, \alpha_2 := 9\beta_2^2$ and $\psi_m$: 

\begin{proposition} \label{modfunct:prop}
	Let $f$ satisfy the nearest neighbor overlapping condition~(\ref{nearestneighbor:eq}). 
	Let $\lambda_\pm:=(\alpha_1 \pm \sqrt{\alpha_1^2 + 4\alpha_2})/2$. Let 
	$C_N$ be the norm squared of $\Phi_N$, considered as a function 
	in $L^2\bigl( (\RR\times[0,2\pi ])^N\bigr)$, and $\rho_N(z)$ the one-particle density. 
	Then 
	\begin{align}
		C_N &= \frac{\lambda_+^{N+1} - \lambda_-^{N+1}}{\lambda_+ - \lambda_-}
				\label{modfunct:eq1} \\
		\begin{split}
		 \rho_N(z)&=\sum_{k=0}^{N-1} \frac{C_k \alpha_1 C_{N-k-1}}{C_N} |\psi_{3k}(z)|^2\\
			&\quad \quad+ \sum_{k=0}^{N-2} \frac{C_k \alpha_2 C_{N-k-2}}{C_N} 
			\bigl(|\psi_{3k+1}(z)|^2 + |\psi_{3k+2}(z)|^2\bigr). \label{modfunct:eq2}
		\end{split}
	\end{align}
\end{proposition}

Before we turn to the proof of the proposition, let us observe that 
it has the periodicity of the one-particle density in the thermodynamic limit as a simple 
consequence. To see this, observe that translating a function $\psi_m$ by a multiple 
of $3$ along the cylinder axis amounts, up to a phase factor, to increasing 
the index $m$ by a multiple of $3$: 
\begin{equation}\label{covarsolv:eq}
	\exp(\ii 3k  y)\psi_m(z- 3k) = \psi_{m+3k}(z).
\end{equation}
The density $\rho_N(z)$ of Eq. (\ref{modfunct:eq2}) is a function of $x = \Re z$ alone 
and vanishes when $x$ is far outside the interval $[0,3N]$. We are interested in the 
limit of the density around the middle of this interval. For a given $z$, only finitely 
many $k$'s contribute to the sum~(\ref{modfunct:eq2}). When $x$ is near the middle of the cylinder, 
the $k$'s that contribute are of the order of $N/2$. But now, because of Eq.~(\ref{modfunct:eq1}),
\begin{equation*}
	\frac{C_k \alpha_1 C_{N-k-1}}{C_N} 
		\sim \frac{1}{\lambda_+-\lambda_-} \frac{\lambda_+^{k+1} \alpha_1 \lambda_+^{N-k}}{\lambda_+^{N+1}}
		 = \frac{\alpha_1}{\lambda_+-\lambda_-}
\end{equation*}
when $k$ and $N-k$ both go to infinity. A similar statement holds for the
coefficient in the second line of Eq.~(\ref{modfunct:eq2}). Hence in the limit $N\rightarrow \infty$, 
the shifted density $\rho_N(z-3 \lfloor N/2 \rfloor)$ converges to 
\begin{equation*}
	 \sum_{k=-\infty}^\infty \Bigl(
	\frac{\alpha_1}{\lambda_+-\lambda_-} |\psi_{3k}(z)|^2 + 
		\frac{\alpha_2 \lambda_+^{-1}}{\lambda_+-\lambda_-} \bigl( |\psi_{3k+1}(z)|^2 + 
			|\psi_{3k+2}(z)|^2\bigr) \Bigr).
\end{equation*}
Because of the covariance~(\ref{covarsolv:eq}), this function is $3$-periodic.
Thus we have obtained the following corollary from 
Prop. \ref{modfunct:prop}:

\begin{corollary}\label{thermolim:cor}
	There is a $3 $-periodic function $\rho(x)$ of the coordinate $x$ alone such that 
	\begin{equation*}
		\lim_{N\rightarrow \infty} \rho_N(z - 3 \lfloor N/2 \rfloor  ) =\rho(x)
	\end{equation*}
	for all $z= x+ \ii y\in\CC$. 	
\end{corollary}

\begin{rem} \label{rem:constdens}
	The corollary does not exclude periods smaller than $3 $. 
	The density can even be constant. For example, when $f$ is a step function 
	taking only two values - one on the intersection of $\supp f$ and the shifted supports 
	$\supp f(\cdot -3n )$ and another one on the remainder of $\supp f$ -  
	a suitable choice of the two values gives a constant density. However, for most 
	choices, the density will have a non-trivial periodicity. 
\end{rem}

Now let us turn to the proof of Prop. \ref{modfunct:prop}.
The proof of the proposition rests on a particular representation of $\Phi_N$ 
in terms of ``monomer'' and ``dimer'' functions 
\begin{equation} \label{defpolyfunct:eq}
   \begin{aligned}
 	u_{\{k\}}(z)&:=\sqrt{\alpha_1} \psi_{3k}(z),\\
	u_{\{k,k+1\}}(z_1,z_2)&:= 
			 - \sqrt{\alpha_2} \bigl(\psi_{3k+1} \wedge \psi_{3k+2}\bigr)(z_1,z_2)
   \end{aligned}
\end{equation}
where $k\in\ZZ$. 
In the following lemma, the modified wave function $\Phi_N$ is written as a sum over 
partitions $X_1,..,X_D$ of $\{0,...,N-1\}$ into monomers $\{k\}$ and dimers $\{k,k+1\}$. 
We will always assume that the partitions are \emph{ordered}, i.e., the elements of $X_1$ are 
smaller than those of $X_2$, etc. 

\begin{lemma} \label{solvpoly:lem}
	The modified wave function $\Phi_N$ can be represented as a sum over 
	ordered monomer-dimer partitions of $\{0,..,N-1\}$:
	\begin{equation} \label{solvpoly:eq}
		\Phi_N = \sum_{(X_1,..,X_D)} u_{X_1}\wedge .. \wedge u_{X_D}.
	\end{equation}
\end{lemma}

\begin{proof}
	The function $\Phi_N$ is defined as the power of a determinant. This leads to the expression
	\begin{equation*}
		\Phi_N(z_1,..,z_N) = \frac{1}{\sqrt{N!}} 
			\Bigl( \sum_{\pi}\sgn(\pi)
		 \phi_{\pi(0)}(z_1)... \phi_{\pi(N-1)} (z_N)\Bigr)^3.
	\end{equation*}
	The sum is over permutations of $\{0,\dots,N-1\}$.
	Expanding the power of the sum, we see that $\Phi_N$ equals $\sqrt{N!}$ times 
	the antisymmetrization of 
	\begin{equation} \label{eq:sita}
		\sum_{\sigma, \tau} \sgn(\sigma \tau) 
		\prod_{j=1}^N \phi_{j-1}(z_j) \phi_{\sigma(j-1)}(z_j)
                \phi_{\tau(j-1)}(z_j).
	\end{equation}	
	Suppose $\sigma,\tau$ give a non-vanishing 
	contribution to the sum above. Then because of
	 the nearest neighbor overlapping condition, they must satisfy 
	\begin{equation} \label{nnpermut:eq}
		\forall k\in \{0,..,N-1\}:\ |\sigma(k)-k|\leq 1,\ |\tau(k)-k|\leq 1,\ |\tau(k)-\sigma(k)|\leq 1.
	\end{equation}
	As a consequence, $\sigma$ and $\tau$ are products of disjoint nearest neighbor 
	transpositions $T_1,..,T_r$. The set of transpositions is uniquely determined by 
	the permutations $\sigma, \tau$. It can be represented by a monomer-dimer partition of 
	$\{0,...,N-1\}$, with $r$ dimers corresponding to the transpositions' supports. 
	Thus to each pair of permutations fulfilling Eq.~(\ref{nnpermut:eq}) we can assign 
	a monomer-dimer partition. 
	
	The proof of Eq.~(\ref{solvpoly:eq}) is concluded by the following observation: if in 
	Eq.~(\ref{eq:sita}) we sum over permutations $\sigma, \tau$ that give the \emph{same}
	partition $X_1,...,X_D$, and then antisymmetrize the resulting sum, we obtain 
	$u_{X_1}\wedge.. \wedge u_{X_D}$. 

	To see this, we consider the example $N=3$, $X_1 = \{0\}$, $X_2 = \{1,2\}$ and leave the general case
	to the reader. There are three permutation pairs
	 $(\sigma, \tau)$ giving rise to the partition 
	 $X_1,X_2$, namely all combinations of the identity $\idty$ and the transposition $(1\ 2)$
	except $\sigma=\tau =\idty$. 
	Their contribution to the sum~(\ref{eq:sita}) is 
	\begin{equation*}
	 	\alpha_1 ^{1/2} \beta_2 \psi_0(z_1)\bigl[ -2 \psi_4(z_2)\psi_5(z_3)+ \psi_5(z_2)\psi_4(z_3)\bigr]
	\end{equation*}
	Antisymmetrization and multiplication by $\sqrt{3!}$ gives $u_{\{0\}}\wedge u_{\{1,2\}}$. \qed
\end{proof}

Lemma~\ref{solvpoly:lem} allows us to prove Prop.~\ref{modfunct:prop}, using the orthogonality 
of contributions from different monomer-dimer partitions. 

\begin{theopargself}
\begin{proof}[of Prop. \ref{modfunct:prop}]
	The key observation is that each monomer-dimer partition $X_1,..,X_D$ is associated 
	with a unique set of $y$-momenta $m_1,..,m_N$: by the definition~(\ref{defpolyfunct:eq})
	of $u_X$, we have 
	\begin{equation}\label{eq:partym}
	 	u_{X_1}\wedge..\wedge u_{X_D} = (-1)^{N-D} (\alpha_{N(X_1)}\cdot ..\cdot \alpha_{N(X_D)})^{1/2}
			\psi_{m_1}\wedge ..\wedge \psi_{m_N}.
	\end{equation}
	The $y$ momenta $\vect{m}=(m_1,..,m_N)$ are obtained as follows: when $X_1,..,X_D$ 
	is a monomer partition, $\vect{m}$ equals $(0,3,..,3N-3)$. The vector belonging to 
	a partition containing the dimer $\{k,k+1\}$ is obtained 
	from this reference vector by replacing $3k, 3k+3$ with $3k+1,3k+2$, see Fig.~\ref{fig:mondim}.
	\begin{figure}[here]
 		\resizebox{8cm}{!}{\input{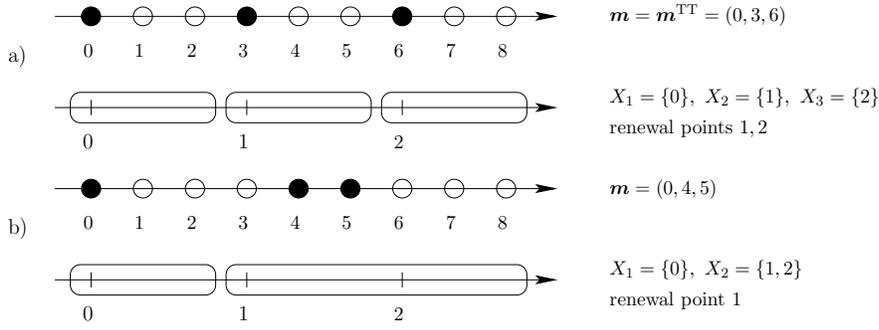}}
        \caption{\label{fig:mondim}
		\small From partitions of $\{0,..,N-1\}$ to wedge products 
		$\psi_{m_1}\wedge..\wedge \psi_{m_N}$: examples for $N=3$.
		a) A monomer partition corresponds to the Tao-Thouless configuration.
		b) When a dimer $\{k,k+1\}$ occurs, $3k,3k+3$ is replaced with $3k+1,3k+2$: 
		here, $k=1$. 
		The notion of renewal points is introduced in Def.~\ref{def:admis} below. 
		They correspond to starting points of rods, $0$ is exluded.
	}
	\end{figure}
	Thus different partitions give rise to different sets of $y$-momenta. 
	whence the orthogonality of the various contributions in Eq.~(\ref{solvpoly:eq}).
	The normalization becomes 
	\begin{equation*} \label{norm:solv:eq}
		C_N = ||\Phi_N||^2 = \sum_{\substack{n_1,..,n_D \in \{1,2\}:\\ n_1+..+n_D =N}} 
			\alpha_{n_1}\cdot.. \cdot \alpha_{n_D}.
	\end{equation*}
	As a consequence, $C_N$ satisfies the recurrence relation 
	\begin{equation} \label{solvrecurr:eq}
		C_N = \alpha_1 C_{N-1} + \alpha_2 C_{N-2}
	\end{equation}
	with the initial conditions $C_0:=1$,  $C_1 = \alpha_1$, whence Eq.~(\ref{modfunct:eq1}). 
	Eqs.~(\ref{solvpoly:eq}) and~(\ref{eq:partym}) lead to 
	\begin{equation*}
		\rho_N (z)= C_N^{-1} \sum_{X_1,..,X_D} \alpha_{N(X_1)}\cdot... \cdot 
		\alpha_{N(X_D)} \sum_{k=1}^D 
			v_{X_k}(z)
	\end{equation*}
	with $v_{\{k\}}(z)=|\psi_{3k}(z)|^2$ and 
	$v_{\{k,k+1\}}(z)=|\psi_{3k+1}(z)|^2+|\psi_{3k+2}(z)|^2$. 
	Changing the order of summation and combining with Eq.~(\ref{norm:solv:eq}) gives Eq.~(\ref{modfunct:eq2}).
	\qed
\end{proof}
\end{theopargself}

\noindent \emph{Remarks. 1.} From Eq.~(\ref{norm:solv:eq})  we see that $C_N$ is 
a monomer-dimer-partition function for a linear chain of length $N$. The recurrence 
relation~(\ref{solvrecurr:eq}) was already given in \cite{heilieb}.

\emph{2.} Similar results hold for $p\neq 3$ and slightly different definitions of 
monomer and dimer functions. 

\emph{3.} If we relax from the nearest neighbor overlapping condition~(\ref{nearestneighbor:eq}), 
Lemma~\ref{solvpoly:lem} still holds true provided we allow polymers of arbitrary length instead 
of only monomers and dimers. The normalization constants then satisfy a recurrence relation of 
infinite order. 

\emph{4.} One last remark concerns the comparison with the torus wave functions introduced 
by Haldane and Rezayi \cite{hr85}. We can define the modified torus wave function $\tilde \Phi_N$ 
by Eq.~(\ref{modfunct:eq}), replacing the functions $\phi_k$ on the right-hand side by periodified 
functions
\begin{equation*}
	\tilde \phi_k(z) := \sum_{n=-\infty}^\infty \exp(\ii n N y) \phi_k(z-3N n) = \sum_{n=-\infty}^\infty 
			\phi_{k+nN}.
\end{equation*}
If in Eq.~(\ref{rootbasefunct-mod:eq}) 
$\phi_k$ is defined with Gaussians $f(x) = \exp(-x^2/2p)$ instead of a function with compact 
supports,
the modified function $\tilde \Phi_N$ is a Haldane-Rezayi torus function. 
When the nearest neighbor overlapping condition is satisfied, the torus function $\tilde \Phi_N$ 
has an explicitly computable normalization and one-particle density. It is associated with 
a monomer-dimer system on a ring, with one long additional polymer covering the whole ring. 
In the limit of long tori, we recover the density of the cylinder modified function. 
Thus for the solvable model, one can check that torus and cylinder functions are equivalent. \\

To summarize, we have seen how the representation~(\ref{solvpoly:eq}) of $\Phi_N$ in terms of a ``quantum monomer-dimer'' system leads to simple formulas for the normalization constants and the one-particle density; 
the periodicity of the one-particle density in the thermodynamic limits stems from the translational 
invariance of the corresponding monomer-dimer system. 

In the next section, we will see that Laughlin's wave function admits a similar representation 
in terms of a polymer system, and the mechanism leading to a periodic one-particle density is 
essentially the same. However, the situation is complicated by the existence of 
polymers of arbitrary length and non-vanishing activity.

\section{Laughlin's cylinder function}

This section is devoted to the proof of the main results of this paper.
In the first subsection, we prove general properties of Laughlin's function. 
We proceed with a polymer representation (Subsect.~\ref{poly:sec}), leading 
to a recurrence relation of infinite order for the normalization constants. This 
relation is exploited in Subsect.~\ref{norm:sec} where we prove an important 
technical result on the asymptotics of normalization constants. 
The existence of the thermodynamic limit of Laughlin's state and its periodicity 
are shown in Subsect.~\ref{thermo:sec}. Symmetry breaking and clustering properties 
are proved in Subsect.~\ref{symclu:sec}.

Throughout this section, we will assume that $p$ is odd. However, most results hold 
for even $p$, with minor modifications (e.g., replacing wedge products with symmetric tensor 
products). The multiplicative constant in the definition~(\ref{laufunct:eq}) of $\Psi_N$ will be fixed as 
\begin{equation} \label{multconst:eq}
	\kappa_N = \frac{1}{\sqrt{N!}} \frac{1}{(2\pi \gamma^{-1} \sqrt{\pi})^{N/2}} 
		\exp(-\frac{1}{2} p^2 \gamma ^2 
		\sum_{j=0}^{N-1} j^2 ).
\end{equation}


\subsection{Basic properties}\label{model:sec}

Laughlin's wave function has a few number of simple, but important properties that we present 
in this subsection. They allow us to derive symmetries of Laughlin's state 
and to show that in the limit of infinitely many particles, Laughlin's state 
cannot give rise to a constant density. This holds regardless of the size of 
the radius,
 in contrast with the results proved in Subsects.~\ref{thermo:sec} and~\ref{symclu:sec}.

All properties rely on the expansion of $\Psi_N$ as a sum of wedge products of 
lowest Landau level basis functions 
\begin{equation*}
	\psi_k(z):= {1\over \sqrt{2\pi \gamma^{-1} \sqrt{\pi}}}\exp(\ii k\gamma y)
	 \exp\bigl(-{1\over 2} (x-k\gamma )^2\bigr), \quad k\in \ZZ.
\end{equation*}
They form an orthonormal set in $L^2(\RR\times[0,2\pi \gamma^{-1}])$. 
One basis function can be transformed into another by a shift in the axial direction: 
\begin{equation*}
	t(\gamma \ex) \psi_k = \psi_{k+1}\quad (k\in \ZZ),
\end{equation*}
where $t(\gamma\ex)$ is the \emph{magnetic translation} $\psi(z) \to \exp(\i\gamma y)\psi(z-\gamma)$. 

The expansion of $\Psi_N$ can be obtained from the expansion 
of the $p$-th power of the Vandermonde determinant into monomials. Define coefficients 
$a_N(\vect{m})$ and $b_N(\vect{m})$ by
\begin{align}
	\Psi_N & = \sum_{0\leq m_1<..<m_N\leq pN-p} a_N(m_1,..,m_N) \psi_{m_1}\wedge ..\wedge \psi_{m_N} 
		\label{lauexp:eq}\\
	\prod_{1\leq j<k\leq N} (z_k-z_j)^p &= \sum_{0\leq m_1,..,m_N \leq pN-p} b_N(m_1,...,m_N) z_1^{m_1}...
		z_N^{m_N}. \label{vandermonde:eq}
\end{align}
A straight-forward computation then gives 
\begin{equation} \label{abrelation:eq}
	a_N(m_1,..,m_N) = b_N(m_1,..,m_N) \exp\bigl(\half \gamma^2\sum_{j=1}^N (m_j^2 - p^2(j-1)^2)\bigr).
\end{equation}
This equation allows us to translate properties of the $p$-th power of the Vandermonde 
determinant into properties of $\Psi_N$. 

The power of the Vandermonde determinant~(\ref{vandermonde:eq})
is a homogeneous polynomial of total degree $pN(N-1)/2$. Therefore $\Psi_N$ has 
definite $y$-momentum
\begin{equation} \label{yinvariance:eq}
	-\ii \sum_{k=1}^N \frac{\partial}{\partial y_k} \Psi_N = \frac{pN(N-1)\gamma}{2}\Psi_N.
\end{equation}
As a consequence, the one-particle density will be independent of the angular coordinate $y$. 
An additional symmetry comes from the 
 ``reversal invariance'' \cite{d93,fgil94}
$	b_N(m_1,..,m_N) = b_N(pN-p -m_N,..,pN-p-m_1):$
Laughlin's state is invariant with respect to 
a $180^\circ$ rotation around the middle of the cylinder, 
\begin{equation} \label{reversalinv}
	s_{p(N-1)\gamma/2}^{\otimes N} \Psi_N = \Psi_N.
\end{equation}
Here $s_a$ refers to a ``magnetic'' rotation, $\psi(z)\to \exp(\ii 2a y) \psi(2a-z)$.
Notice that $s_{r\gamma /2} \psi_k = \psi_{2r-k}$.

Because of Eqs.~(\ref{lauexp:eq}) and~(\ref{yinvariance:eq}), the one-particle density 
can be expressed as:
\begin{equation} \label{density:eq}
	\rho_N (z)= \sum_{k=0}^{pN-p} \la c_k^*c_k \ra_N\, |\psi_k(z)|^2 
		\propto \sum_{k=0}^{pN-p}  \la c_k^*c_k \ra_N 
		\exp\bigl(-(x-k\gamma )^2\bigr).
\end{equation}
$\la \cdot \ra_N$ refers to expectation values in the state $\Psi_N/||\Psi_N||$
and $c_k^*$, $c_k$ are the fermionic creation and annihilation operators for the state $\psi_k$, 
e.g., $c_k^* f = \psi_k \wedge f$. Notice that the expectation values $\la c_k^*c_m\ra$, $k\neq m$,
 vanish 
due to the $y$-invariance~(\ref{yinvariance:eq}).

We are interested in limits of $\rho_N(z - p\lfloor N/2\rfloor \gamma)$ as 
$N \rightarrow \infty$. Eq.~(\ref{density:eq}) shows that any limit point 
of the shifted density is a sum of Gaussians. The following lemma deduces 
several statements from this observation.

\begin{lemma} \label{foutrans:lem}
	Let $(n_k)_{k\in \ZZ}$ be a sequence of numbers in $[0,1]$ that does not identically vanish, 
	and $\rho(x):=\sum_{k=-\infty}^\infty n_k |\psi_k(x)|^2$. Then \begin{enumerate}
		\item $\rho(x)$ cannot be a constant. 
		\item If $\rho(x)$ is periodic, every period must be a multiple of $\gamma$. 
		\item $\rho(x)$ is periodic with period $p\gamma$ if and only if $(n_k)$ is 
		periodic with period $p$. 
	\end{enumerate}
\end{lemma}

\begin{proof}
	$\rho = f*\mu$ is the convolution of the function $f(x):=(2\pi \gamma^{-1} \sqrt{\pi})^{-1}\exp(-x^2)$ 
	and the measure $\mu=\sum_{k\in \ZZ} n_k \delta_{k\gamma }$. 
	With $\rho$ and $\mu$ we can associate tempered distributions $T_\rho$ and $T_\mu$:
	\begin{equation*}
		T_\rho \phi:=\int_{-\infty}^\infty \rho(x) \phi(x) \intd x, \quad T_\mu \phi:=\int_\RR \phi \intd \mu.
	\end{equation*}
	Their Fourier transforms ($\hat T\phi:= T\check \phi$) satisfy 
	\begin{equation} \label{fouconv:eq}
		\hat T_\rho = \sqrt{2\pi} \hat f\thinspace  \hat T_\mu,\quad 
		\hat f(k) = \frac{\gamma}{2\pi } \exp(-k^2 /4). 
	\end{equation}
	The product $\hat f \hat T_\mu$ is the distribution $(\hat f\hat T_\mu) \phi:= \hat
	T_\mu(\hat f\phi)$. 	
	It follows from Eq.~(\ref{fouconv:eq})
	 that the mapping $\mu \mapsto f* \mu$ is injective. Therefore $f$ and $\mu$ 
	must have the same periodicity. Since any period of $\mu$ is obviously a multiple of $\gamma $, 
	the same must be true for periods of $\rho$. This excludes a constant density, since constant 
	functions admit periods that are not multiples of $\gamma$.  \qed
\end{proof}

Notice that it is not completely trivial that the density 
cannot be constant. If instead of Gaussians we added up more general functions, we could 
obtain a constant. Examples include functions of compact support, see the remark on 
p.~\pageref{rem:constdens}, and log-concave functions, as was shown in \cite{bl75}, Sect.~1.4.3.

Observe also that there is a relationship not only between the periods 
of $(n_k)$ and $\rho(x)$, but also between the amplitudes of oscillation.
If $\rho(x)$ is $p\gamma$-periodic, the $k$-th Fourier coefficients can be expressed in terms 
of occupation numbers:
\begin{equation*}
 	\frac{1}{p\gamma}\int_0^{p\gamma} \rho(x)\exp(-\ii \frac{2\pi k x}{p\gamma})\intd x 
	= \frac{1}{p}\frac{1}{2\pi} \exp(-\frac{\pi^2 k^2}{p^2\gamma^2})
	\sum_{j=0}^{p-1}n_j \exp(-\ii \frac{2\pi kj}{p}).
\end{equation*}
This relation generalizes Poisson's summation formula, which shows that the 
Fourier coefficients of $\rho(x)$ in the filled Landau level ($p=1$ and $n_k \equiv 1$) 
are $(2\pi)^{-1} \exp( - \pi^2 k^2/\gamma^2)$. 


\subsection{Associated polymer system and renewal equation}\label{poly:sec}

In this subsection we show that Laughlin's wave function admits a representation
as a sum over certain partitions of $\{0,..,N-1\}$, analogous to Lemma~\ref{solvpoly:lem} 
for the solvable model. 
Instead of only monomers or dimers, the partition may contain longer polymers or ``rods''
$X = \{j,..,j+n-1\}$. The normalization becomes a discrete polymer partition function 
\cite{gruberkunz}
with translationally invariant activity and satisfies a discrete renewal equation, 
as explained at the end of this subsection. 

Let $N(X)$ denote the  cardinality or ``length'' of a rod $X$, and let
 $\mathcal{P}_N$ be the set of ordered partitions $X_1,..,X_D$ of 
$\{0,1,..,N-1\}$ into rods $X_j$. The following holds:
%
%

\begin{proposition}[Associated polymer system]\label{polymer:prop}
 	There is a mapping associating with
	each rod $X$ an antisymmetric function $u_X$ of  $N(X)$ complex variables so that 
	\begin{align}
		\Psi_N& = \sum_{(X_1,..,X_D) \in \mathcal{P}_N} u_{X_1}\wedge..\wedge u_{X_D},
			\label{polymer-function} \\
		C_N:=||\Psi_N||^2&=\sum_{(X_1,..,X_D)\in \mathcal{P}_N} \Phi(X_1)\cdot..\cdot \Phi(X_D), \qquad 
			\Phi(X)=||u_X||^2\label{polymer:eq2}.
	\end{align}
\end{proposition}

The expansion~(\ref{polymer-function}) for $N=3$ particles and $p=3$ 
is explicitly written down 
in~\cite{jls}, Eq.~(10), 
 up to a small difference: in \cite{jls}, the sum is over partitions of the set 
$\{-1,...,3N-2\}$ instead of $\{0,...,N-1\}$, and rods $X$ have length $|X|=3,6,...$ instead 
of $|X|=1,2,...$ as is the case here. 

We will see that the family of ``polymer functions'' ($u_X$) has two important additional 
properties. 
The first is a kind of localization: $u_X$ is a sum of wedge products of 
functions $\psi_k$ with indices $k$ in $\{p\min X,...,p\max X -p\}$. The second 
property is translational covariance:
\begin{equation}\label{covar:eq}
	\forall j\in \ZZ,\qquad u_{j+X} = t(j p\gamma \ex)^{\otimes N(X)}u_X.
\end{equation}
Thus the shift of a rod results in the magnetic translation of the corresponding function. 
This covariance is at the origin of $p\gamma$-periodicity in Laughlin's state. 

Eq.~(\ref{polymer:eq2}) says that the normalization $C_N$ is a polymer partition function 
with activity $\Phi(X)$. As a consequence of the covariance~(\ref{covar:eq}), the activity is translationally 
invariant, i.e., the activity of 
a rod depends on its length $N(X)$ only: 
we can define non-negative numbers $(\alpha_n)_{n\in\NN}$ by\footnote{Here, we use that $||\cdot||$ refers to integration 
over the \emph{infinite} cylinder $\RR\times [0,2\pi \gamma^{-1}]$. On finite cylinders, 
the activity is not completely translationally invariant due to boundary effects, although 
Prop.~\ref{polymer:prop} and the covariance~(\ref{covar:eq}) stay true.}
\begin{equation}\label{activity:eq}
 	\Phi(X) =||u_X||^2 = \alpha_{N(X)}.
\end{equation}
The numbers $(\alpha_n)$ will appear as coefficients in a recurrence relation for the normalization 
$C_N$.

The proof of the previous proposition makes crucial use of a product rule proved in \cite{fgil94} 
for the expansion coefficients $b_N(\vect{m})$. 
 It is more easily expressed with some auxiliary definitions.

\begin{definition}\label{def:admis}
	Let $\vect{m} =(m_1,..,m_N) \in \ZZ^N$ with $m_1\leq ..\leq m_N$. 
	\begin{itemize}
		\item $\vect{m}$ is \textbf{$N$-admissible} if 
			\begin{equation*}
				\sum_{j=1}^k m_j \geq \sum_{j=1}^k p(j-1) 
			\end{equation*}
			for all $k \in \{1,...,N\}$ with equality for $k=N$. 
		\item $k\in \{1,...,N-1\}$ is a \textbf{renewal point} of $\vect{m}$ if 
			\begin{equation*}
				\sum_{j=1}^k m_j = \sum_{j=1}^k p(j-1).
			\end{equation*}
		\item $\vect{m}$ is \textbf{reducible} if it admits a renewal point $k\in \{1,..,N-1\}$, 
			and \textbf{irreducible} in the opposite case. 
	\end{itemize}
\end{definition}
Note that $N$-admissibility actually implies $0\leq m_1,..,m_N\leq pN-p$. 
The reducibility of a sequence is most easily visualized with  the help of
the sequence of occupation numbers in Laughlins' state, see~\cite{jls}, Sect.~4.1.

There are various equivalent characterizations of $N$-admissibility 
\cite{fgil94,rh94,ktw01}. In all of them, the so-called Tao-Thouless 
configuration \cite{tt83}
\begin{equation} \label{taothoum}
	\taothou:=(0,p,2p,..,pN-p)
\end{equation}
plays the role of a reference configuration. In the proof of the next lemma, 
we will use the following: a vector $\vect{m}$ is $N$-admissible 
if and only if it is \emph{majorized} by $\taothou$, which is equivalent to the existence 
of a doubly stochastic matrix $P$ such that 
\begin{equation} \label{mpt}
	\vect{m} = P\taothou,
\end{equation}
see e.g. \cite{hlp}, p.~49. 

This characterization allows us to visualize 
renewal points as block-diagonal matrices:
If $\vect{m}$ is given by Eq,~(\ref{mpt}) for some block-diagonal, doubly stochastic matrix 
$P$ that has an upper left $k\times k$ block, $k$ is a renewal point of $\vect{m}$. Conversely, 
if $k$ is a renewal point of the $N$-admissible vector $\vect{m}$, every doubly stochastic matrix 
fulfilling Eq.~(\ref{mpt}) must be block-diagonal. 

By the Birkhoff-von Neumann theorem, a matrix is doubly stochastic if and only if 
it is a convex combination of permutation matrices. Interestingly, the only matrices 
that we shall need are equally weighted averages of permutation matrices, 
see Eq.~(\ref{stochmat}) below. 

The following lemma contains fundamental properties of the expansion coefficients 
$a_N(\vect{m})$:

\begin{lemma}  \label{prodrule:lem}
	Let $\vect{m}=(m_1,..,m_N)$ with $m_1\leq ..\leq m_N$.
	\begin{enumerate}
		\item  Suppose
		$a_N(m_1,..,m_N)\neq 0$. Then $\vect{m}$ is $N$-admissible. 
		\item Suppose $k\in\{1,...,N\}$ is a renewal point of $\vect{m}$. Then:
		\begin{enumerate}
			\item $\vect{m}$ is $N$-admissible if and only if $(m_1,..,m_k)$ 
		and $(m_k-pk,...,m_N-pk)$ are $k$- resp. $(N-k)$-admissible. 
			\item The following \textbf{product rule} holds:
		\begin{equation} \label{prodrule:eq}
			a_N(m_1,..,m_N) = a_k(m_1,..,m_k)a_{N-k}(m_{k+1}-pk,..,m_N-pk).
		\end{equation}
		\end{enumerate}
	\end{enumerate}
\end{lemma}
We will frequently write Eq.~(\ref{prodrule:eq}) as
\begin{equation*}
	a_N(\vect{m}) = a_N(\vect{m}_1^N) = a_k(\vect{m}_1^k) a_{N-k}(\vect{m}_{k+1}^N - pk)
\end{equation*}
using the short-hands $\vect{m}_a^b:= (m_a,..,m_b)$ and $(x_1 - b,..,x_n-b) =: \vect{x}-b$. 

\begin{proof}
	The previous lemma holds true if $a_N$ is replaced with the coefficients $b_N$ defined in 
	Eq.~(\ref{vandermonde:eq}). The corresponding statements for $b_N$ have been proved in 
	\cite{fgil94}, and the statements for $a_N$ may be deduced with the help 
	of Eq.~(\ref{abrelation:eq}).
	
	The proof given in \cite{fgil94}
	 uses the polynomial character of the Vandermonde determinant. 
	However, in view of the results on the solvable model (Sect. \ref{solv:sec}), we should like 
	to stress the importance of the fact that we have the power of a determinant. Therefore 
	we give a new proof. For the sake of 
	simplicity, we take $p=3$; the proof for other values of $p$ is similar. 

	\emph{1.} Expanding 
	the right-hand side of 
	\begin{equation} \label{bnsigns:eq}
		\prod_{1\leq j<k\leq N} (z_k-z_j)^3 
		= \left( \sum_{\pi \in \perm_N} \sgn(\pi) z_1^{\pi(1)-1}...z_N^{\pi(N) - 1} 
						\right)^3
	\end{equation}
	we find 
	\begin{equation} \label{sumbN}
		b_N(m_1,..,m_N) = \sum_{\substack{\pi, \sigma, \tau \in \perm_N: \\ 
		\vect{m} = \pi + \sigma + \tau -3}} \sgn(\pi\sigma \tau),
	\end{equation}
	where  the short-hand $\vect{m} = \pi + \sigma + \tau -3$ in the subscript stands for
	\begin{equation*}
		\forall j\in \{1,..,N\}:\ m_j = \pi(j)+ \sigma(j)+ \tau(j) - 3.
	\end{equation*}
	Now suppose $b_N(\vect{m}) \neq 0$. Then there exist permutations $\pi, \sigma, \tau$ adding 
	up to $\vect{m}$ in the sense of the previous equation. 
	In terms of the associated permutation matrices,
	\begin{equation} \label{stochmat} 
		\vect{m}= P\taothou,\quad P=\frac{1}{3}(P_\pi + P_\sigma+ P_\tau)
	\end{equation}
	where $\taothou$ is as in Eq.~(\ref{taothoum}) with $p=3$.
	$P$ is doubly stochastic and therefore $\vect{m}$ is $N$-admissible. 
	Since by Eq.~(\ref{abrelation:eq}) $a_N(\vect{m})\neq 0$ if and only if $b_N(\vect{m}) \neq 0$, the first statement of 
	the lemma follows. 

	\emph{2.} Let $\vect{m}= (m_1,...,m_N)$ with $m_1\leq ..\leq m_N$. 
		and let $k$ be a renewal point of $\vect{m}$. 
	\emph{2.(a)} is proven in \cite{ktw01}. For \emph{2.(b)}, suppose $\pi, \sigma, \tau$ are 
	permutations 	such that Eq.~(\ref{stochmat}) holds. Then the matrix $P$ from this equation  
	must be block-diagonal with an upper left 
	$k\times k$ block. This means that $\pi,\sigma, \tau$ 
	must leave $\{1,..,k\}$ and $\{k+1,..,N\}$ invariant. 
	As a consequence, the sum~(\ref{sumbN}) factorizes into two sums
	\begin{align*}
		b_N(\vect{m})&= \Bigl( \sideset{}{'}\sum_{\pi', \sigma', \tau'}
				 \sgn(\pi'\sigma'\tau') \Bigr)
			 \Bigl( \sideset{}{''}\sum_{\pi'', \sigma'', \tau''}
				 \sgn(\pi''\sigma''\tau'') \Bigl).
	\end{align*}
	The first sum is over permutations of $\{1,..,k\}$ such that 
	\begin{equation*}
		\vect{m}_1^k= \pi'+\sigma'+\tau'-3,
	\end{equation*}
	and equals $b_k(m_1,..,m_k)$.
	The second sum is over permutations of $\{k+1,..,N\}$ 
	such that 
	\begin{equation*}
		\vect{m}_{k+1}^N= \pi''+\sigma''+\tau''-3.
	\end{equation*}
	Rewriting this equation with 
	permutations of $\{1,..,N-k\}$ instead of $\{k+1,..,N\}$, one finds that the 
	second sum  
	equals $b_{N-k}(\vect{m}_{k+1}^N-3k)$. The product rule thus holds for $b_N(\vect{m})$. 
	A straight-forward computation on the exponential factor in~(\ref{abrelation:eq}) 
	allows us to conclude that 
	$a_N(\vect{m})$ satisfies the product rule too. \qed
\end{proof}

Now we are able to prove Prop.~\ref{polymer:prop}.

\begin{theopargself}
\begin{proof}[of Prop. \ref{polymer:prop}]
	In the expansion~(\ref{lauexp:eq}), we group together admissible $\vect{m}$'s
	that have the same set of renewal points $r_1<..<r_D$. 
	The product rule~(\ref{prodrule:eq}) gives 	
	\begin{multline} \label{wfprod:eq}
		a_N(\vect{m}) \psi_{m_1}\wedge ..\wedge \psi_{m_N} \\
		=\Bigl(a_{r_1}(\vect{m}_1^{r_1})\psi_{m_1}\wedge..\wedge \psi_{m_{r_1}}\Bigr)
			\wedge
			\Bigl( a_{r_2-r_1}(\vect{m}_{r_1+1}^{r_2} - p r_1) 
				\psi_{m_{r_1+1}}\wedge ..\wedge \psi_{m_{r_2}}\Bigr)\\
			 \wedge .. \wedge 
			\Bigl( a_{N-r_D}(\vect{m}_{r_D+1}^{N} - pr_D) \psi_{m_{r_D+1}}\wedge ..\wedge \psi_{m_N} \Bigr).
	\end{multline}
	The renewal points split $\vect{m}$ into $D+1$ irreducible blocks. 
	This motivates the definition of polymer functions $u_X$ as sums over 
	irreducible sequences:
	\begin{equation} \label{defpolyfunctions:eq}
		u_{\{a,..,a+n-1\}}:=\sum_{\substack{\vect{m}\ n-\text{admissible}\\ \text{and irreducible}}}
		 a_n(\vect{m})\, \psi_{m_1+pa}\wedge ..\wedge \psi_{m_n+pa}.
	\end{equation}
	These functions fulfill 
	the covariance~(\ref{covar:eq}).  
	Combining the expansion~(\ref{lauexp:eq}) and Eq.~(\ref{wfprod:eq}), we obtain 
	\begin{equation*}
		\Psi_N = \sum_{D=0}^{N-1}
		 \sum_{0<r_1<..<r_D<N} u_{\{0,..,r_1-1\}}\wedge u_{\{r_1,..,r_2-1\}}
			\wedge .. \wedge u_{\{r_D,..,N-1\}},
	\end{equation*}
	which concludes the proof. \qed
\end{proof}
\end{theopargself}


The results presented in this article rely heavily on the 
relationship of 
$\Psi_N$ and polymer systems presented in Prop.~\ref{polymer:prop}. Let us recall 
 some basic facts 
on polymer systems with translationally invariant activity supported by rods $\{j,..,j+n-1\}$ 
and the link to renewal equations. 
 In view of Eq.~(\ref{activity:eq}), the expression~(\ref{polymer:eq2}) 
for the polymer partition function $C_N$ becomes
\begin{equation} \label{normalization:eq}
	C_N = \sum_{\substack{D, n_1,..,n_D\in \NN:\\ n_1+..+n_D = N}} \alpha_{n_1}...\alpha_{n_D}.
\end{equation}
It follows that $C_{N+M}\geq C_N C_M$. This supermultiplicativity 
is a general property of polymer partition functions, see \cite{gruberkunz}.
 As a consequence, we can define 
\begin{equation} \label{pressure:eq}
	-\ln r:= \sup_{N\in \NN} \frac{1}{N} \ln C_N = \lim_{N\rightarrow \infty} \frac{1}{N} \ln C_N,
\end{equation}
where we use the convention $-\ln 0 = \infty$. The left-hand side $-\ln r$ is the \emph{pressure} 
of the polymer system; note that $r$ is the radius of convergence of the power series $\sum_n C_n t^n$. 
 If we suppose in addition that there exists some $\xi>0$ such that 
the rescaled activity $\xi^N \Phi: X\mapsto \xi^{N(X)} \Phi(X)$ is \emph{stable}, i.e., 
\begin{equation} \label{stability:ineq}
	\forall k\in \ZZ:\ \sum_{X\ni k} \frac{\xi^{N(X)}\Phi(X)}{N(X)}<\infty,
\end{equation}
the pressure $-\ln r$ is finite. This follows from general theorems \cite{gruberkunz} but can be shown 
here using a particularly 
simple argument: the stability condition~(\ref{stability:ineq}) is equivalent to $\sum_n \xi^n \alpha_n<\infty$. 
It is fulfilled for some $\xi>0$ if and only if the power series $\sum_n \alpha_n t^n$ has a nonvanishing radius 
of convergence. On the other hand, Eq.~(\ref{normalization:eq}) leads to the formal power series identity 
\begin{equation} \label{powerseries:eq}
	C(t)=1+\sum_{n=1}^\infty C_n t^n = \frac{1}{1-\sum_{n=1}^\infty \alpha_n t^n}= \frac{1}{1-\alpha(t)}.
\end{equation}
Therefore, if $\alpha(t)$ has a positive radius of convergence, so has $\sum_n C_n t^n$, whence $r>0$ and 
 $-\ln r<\infty$. 

The type of polymer systems envisaged here bears an interesting connection to renewal theory, as observed
 by \cite{ivz} (see \cite{feller} for an account on renewal theory). The formal power series
 identity~(\ref{powerseries:eq})
is equivalent to the recurrence relation
\begin{equation} \label{renewal:eq}
	\forall n\in \NN:\quad C_n= \alpha_1 C_{n-1}+...+ \alpha_n, \quad C_0:=1.
\end{equation}
This recurrence relation is known in stochastics as a (discrete) \emph{renewal equation}. 
If we interpret $\ZZ$ as a discrete time axis and starting points of rods as events, 
the activity $(\alpha_n)$ is related to a probability distribution on waiting times between events,  
and $C_n$ is related to the probability that an event occurs (i.e., that $n$ is a renewal point). More
precisely, we suppose that $r>0$ and set $p_n:= r^n\alpha_n$ and $u_n:=r^n C_n$. \label{assocrenew:page}
Suppose that the sequence of activities is \emph{aperiodic}, i.e.,
\begin{equation} \label{aperiodicity:eq}
	\gcd\{n\in\NN \mid \alpha_n >0 \} =1.
\end{equation}
Then one of the following cases necessarily holds: 
\begin{enumerate}
\item $(p_n)$ defines a probability measure on $\NN$ with finite mean $\mu$. In this case 
	$u_n =r^n C_n \rightarrow  \mu^{-1} > 0$. (The associated renewal process is positive recurrent.)
\item $(p_n)$ defines a probability measure on $\NN$ with infinite mean $\mu=\infty$. Then 
	$u_n = r^n C_n \rightarrow 0$ (but $\sum_n u_n =\infty$). (The renewal process is null recurrent.)
\item $(p_n)$ defines a defective measure on $\NN$ (i.e.,  $\sum_n p_n<1$). Then 
	$\sum_n u_n = (1-\sum_n p_n)^{-1}<\infty$, whence $u_n \rightarrow 0$. (The renewal process is transient.)
\end{enumerate}
See \cite{feller} for a proof. We will refer to the first case as a renewal process with \emph{finite mean}. 
Thus in any case, $r^n C_n \rightarrow q\geq 0$, but $q>0$ if and only if the associated renewal process 
has finite mean.

\subsection{Large $N$-asymptotics of normalization constants}\label{norm:sec}

In this subsection, we exploit the relation of $\Psi_N$ with polymer systems 
and renewal equation to investigate the asymptotics of the normalization constant $C_N = ||\Psi_N||^2$. 
The result comes in two parts: first, for all values of the radius, there exist a 
$r>0$ and $q\geq 0$ such that $C_N r^N \rightarrow q \geq 0$ (Lemma~\ref{normalization:lem}). 
Second, for sufficiently large $\gamma$ (thin cylinders), $q$ is strictly positive (Theorem~\ref{laga:theo}).
Thus the associated polymer system, suitably rescaled,
has a stable activity in the sense of Eq.~(\ref{stability:ineq}), and on sufficiently 
thin cylinders, the associated renewal process has finite mean. 

In addition, we give lower and upper bounds on $r$ that are interesting in the context of 
Laughlin's plasma analogy. Our bounds are consistent with results by \cite{forr91} 
on the free energy of a jellium system placed on a cylinder with large radius. 

\begin{lemma} \label{normalization:lem}
	Let $p\in \NN$ and $\gamma >0$. There exist $r>0$ and $q\geq 0$ such that 
	\begin{equation*}
		-\ln r= \lim_{N\rightarrow \infty} {1\over N} \ln C_N = \sup_N {1\over N }\ln C_N, 
		\quad q= \lim_{n\rightarrow \infty} C_n r^n.
	\end{equation*}	
	Moreover, $r$ satisfies the bound
	\begin{equation}\label{normaliobo:eq}
		\Bigl( e^p \sum_{n_1,..,n_p \in \ZZ:\atop n_1+..+n_p=0} \exp\bigl(-{\pi^2\over p\gamma ^2} 
		(n_1^2+.. + n_p^2)\bigr)\Bigr)^{-1}
		\leq r\ p^{1-{p\over 2}} ({e\gamma \over \sqrt{\pi}})^{1-p}\leq 1.
	\end{equation}
\end{lemma}

\begin{proof}
	We know already that $-\ln r = \lim_n n^{-1} \log C_n$ exists, see~(\ref{pressure:eq}). 
	The strict positivity of $r$ will follow from Ineq.~(\ref{normaliobo:eq}).
	The observation of the previous 
	subsection yields the existence of $q=\lim_n C_n r^n$; note that $\alpha_1 = ||\psi_0||^2 =1$ 
	so that the sequence of activities $(\alpha_n)$ fulfills the aperiodicity 
	condition (\ref{aperiodicity:eq}). 

	Thus it remains to prove Ineq. (\ref{normaliobo:eq}). 	
	The idea is to use the representation (\ref{powdet:eq}) of $\Psi_N$ as the 
	$p$-th power of a determinant times $1/\sqrt{N!}$ and to give lower and upper bounds on $C_N$ 
	using H\"older's and Hadamard's inequalities.
	These inequalities have already been used in \cite{fgil94}, Sect. 3.3., to derive 
	bounds on the free energy of jellium on a sphere. 
	 We start with the application of 
	Hadamard's inequality, which gives 
	\begin{equation*}
		|\Psi_N(z_1,..,z_N)|^2 \leq \frac{1}{N!} \prod_{j=1}^N \bigl(\sum_{k=0}^{N-1}
		 |\varphi_k(z_j)|^2\bigr)^p.
	\end{equation*}
	It follows that 
	\begin{equation*}
		C_N \leq \frac{1}{N!} \Bigl( \int_{-\infty}^\infty \frac{1}{\sqrt{\pi}} 
			\bigr(\sum_{k=0}^{N-1}\exp(-\frac{(s-pk\gamma)^2}{p})\bigl)^p \intd s \Bigr)^N.
	\end{equation*}
	The integral from $-\infty$ to $-p\gamma/2$ and from $(N-1/2)p\gamma$ to $\infty$ 
	can be bounded by an $N$-independent constant. The integral from $-p\gamma/2$ to 
	$(N-1/2)p\gamma$ is bounded from above by 
	\begin{equation}\label{midint:eq}
		 \frac{1}{\sqrt{\pi}} N \int_0^{p\gamma} f(x)^p \intd x,\quad
		f(x):=\sum_{k=-\infty}^\infty \exp\bigl(-\frac{(x-pk\gamma)^2}{p}\bigr).
	\end{equation}
	Representing $f$ as a Fourier series via Poisson's summation formula, we find 
	that the first expression in~(\ref{midint:eq}) equals $Nb(\gamma)$ with 
	\begin{equation*}
		b(\gamma) = p^{1-\frac{p}{2}} \bigl(\frac{\sqrt{\pi}}{\gamma})^{p-1}\sum_{
		\substack{n_1,..,n_p\in \ZZ:\\ n_1+..+n_p =0}} \exp\bigl(
			-\frac{\pi^2}{p\gamma^2}(n_1^2+..+n_p^2)\bigr).
	\end{equation*}
	 Thus we get
	$C_N \leq(N b(\gamma)+ c)^N/N!$,
	from which the lower bound on $r$ is easily obtained.

	Now we turn to a lower bound for $C_N$.
	With H\"older's inequality written as 
	\begin{equation*}
		\int_\Omega g^p \geq |\Omega|^{p-1}\ \bigl|\int_\Omega g\bigr|^p,
	\end{equation*}
	applied to the domain of integration 
	$([-p\gamma/2, (N-1/2)p\gamma]\times [0,2\pi \gamma^{-1}])^N$,
	we find
	\begin{equation*}
		C_N \geq \Bigl( {N!\sqrt{\pi}^N\over (Np\gamma )^N}\Bigr) ^{p-1} \sqrt{p}^{pN} 
	 		\prod_{k=0}^{N-1}\Bigl(1-\epsilon_{N-k-{1\over 2}} - \epsilon_{k+{1\over 2}}\Bigr)^p
	\end{equation*}
	where $\epsilon_m=[\erfc(m\sqrt{p}\gamma)]/2$ and 
	$\erfc$ 
	is the complementary error function.
	The product over $k$ does not contribute to $\lim N^{-1}\log C_N$. 
	 Making use of Stirling's formula, we obtain the desired upper bound to $r$. \qed
\end{proof}

\begin{rem}
	The bounds~(\ref{normaliobo:eq}) lead to a statement on the thick cylinder asymptotics of $r$:
	$
		r = O(\gamma^{p-1}) \ \text{as}\ \gamma \rightarrow 0.
	$
	This complements the thin cylinder ($\gamma \rightarrow \infty$) asymptotics given in 
	Eq.~(\ref{laga:eq}) below.
\end{rem}

Lemma~\ref{normalization:lem} leaves open the question whether $q>0$ or $q=0$, i.e.,
whether the associated renewal process has finite or infinite mean. In order to answer 
this question, it is useful to have a closer look at the activity $(\alpha_n)$. 

\begin{lemma}[$\gamma$-dependence of the activity] \label{actibo:lem}
	Monomers have activity $\alpha_1=1$. The activity of a polymer of length $N\geq 2$ 
	is a polynomial of $\exp(-\gamma^2)$ with minimal degree $p(N-1)$ and 
	coefficient in $\NN_0$. In particular, 
	\begin{equation*}
		\alpha_N = O\Bigl( \bigl(\exp(-\gamma^2)\bigr)^{p(N-1)} \Bigr) \ \text{as}\ 
		\gamma \rightarrow \infty.
	\end{equation*}
\end{lemma}
Hence, in the thin cylinder limit, only monomers have a non-vanishing activity. 

\begin{proof}
	The monomer functions are $u_{\{k\}}(z)=\psi_{pk}(z)$, whence 
	$\alpha_1=1$. Combining Eqs.~(\ref{abrelation:eq}) and (\ref{defpolyfunctions:eq}), we find
	\begin{equation} \label{alphan-polynomial:eq}
		\alpha_N = \sum_{\vect{m}\ \text{irreducible}} |b_N(\vect{m})|^2 
			(e^{-\gamma^2})^{\sum_{j=1}^N (p^2(j-1)^2-m_j^2)}.
	\end{equation}
	By  Eq.~(\ref{bnsigns:eq}), the expansion coefficients $b_N(\vect{m})$ are sums 
	of signs of permutations and therefore integers. The proof of the lemma is concluded 
	by the following observation: 
	if $m_1\leq...\leq m_N$ is $N$-admissible and irreducible,
	\begin{equation} \label{sumsquares:ineq}
		\sum_{j=1}^N (p^2(j-1)^2-m_j^2) \geq p(N-1).
	\end{equation}
	Due to $N$-admissibility, we can write $m_j = p(j-1) + \nu_j-\nu_{j-1}$ 
	with $\nu_0=\nu_N =0$ and $\nu_1,..,\nu_{N-1}\geq 0$ (see also \cite{fgil94}, 
	Property~3). $\nu_k$ is just $\sum_{j=1}^{k} [m_j - p(j-1)]$. 
	Because of irreducibility, $\nu_1,..,\nu_k$ must be strictly positive. 
	In the left-hand side of~(\ref{sumsquares:ineq}), we insert the expression of $m_j$ in terms of 
	$\nu_k$ and perform a summation by parts, and obtain Ineq.~(\ref{sumsquares:ineq}). \qed
\end{proof}

Now recall that $q>0$ if and only if $\sum \alpha_n r^n =1$ 
and $\sum n \alpha_n r^n <\infty$. 
The crucial observation is that these two conditions are automatically fulfilled when 
the generating series of $(\alpha_n)$ 
has a radius of convergence $R_\alpha$ strictly larger 
than the radius of convergence $r$ of the power series with coefficients $(C_n)$.
 Therefore we are going to compare domains of convergence. 

\begin{figure}[here]
 	\begin{center}
     		\resizebox{6cm}{!}{\input{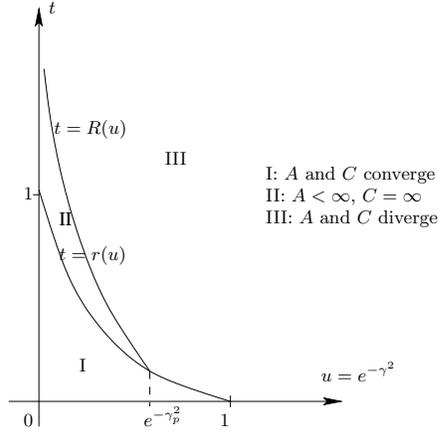}}
        \end{center}
 	\caption{\label{domco_fig}
		\small Domains of convergence of $A(t,u)$ and $C(t,u)$ for $p\geq 2$. The curve
		$R(u)$ delimits the domain of convergence of $A$, $r(u)$ the domain of convergence 
		of $C$. 
		Both series diverge when $u\geq 1$. We know that $r(u)=1+O(u)$ and 
		$R(u) \geq {\rm const}\cdot u^{-p}$ as
		$u\rightarrow 0$. When $u<\exp(-\gamma_p^2)$, $r(u)<R(u)$. It is an open question whether 
		the curves $r$ and $R$ touch for some $u=\exp(-\gamma_p^2)$ strictly below $1$. 
	}
\end{figure}

For a monomer system with $\alpha_1=1$ (and $\alpha_n=0$ for $n \geq 2$), 
 the quantities are trivial to compute:
\begin{equation} \label{monomervalues}
	C_n \equiv1,\quad r=1,\quad R_\alpha = \infty,\quad q=1.
\end{equation}
We will show that on sufficiently thin cylinders, the quantities 
$r,R_\alpha,q$ take values 
close to the monomer values~(\ref{monomervalues}). 
For this purpose it is useful to keep track of the $\gamma$-dependence in the notation. 
By Lemma~\ref{actibo:lem} and Eq.~(\ref{normalization:eq}), 
the activity and the normalization constants are polynomials 
of $e^{-\gamma^2}$ with coefficients in $\NN$. 
Therefore we write $\alpha_n(e^{-\gamma^2})$, $C_n(e^{-\gamma^2})$. 
The power series
\begin{equation*}
 	C(t, e^{-\gamma^2}):= 1+ \sum_{n=1}^\infty C_n(e^{-\gamma^2}) t^n ,\quad 
 	A(t, e^{-\gamma^2}):= t+ \sum_{n=2}^\infty \alpha_n(e^{-\gamma^2}) \tp t^n
\end{equation*}
 are actually power series of \emph{two} variables, $t$ and $u = e^{-\gamma^2}$. They have 
non-negative integer coefficients and are related through 
\begin{equation*}
	C(t,u) = {1\over 1-A(t,u)}, 
\end{equation*}
see Eq.~(\ref{powerseries:eq}).
The curves $r =r (u)$ and $R_\alpha=R_\alpha(u)$ 
delimit the domains of convergence of $C(t,u)$ and $A(t,u)$, see Fig.~\ref{domco_fig}. 

The following theorem states that the curve $r(u)$ stays strictly below $R(u)$, at least for 
small $u$ (large $\gamma$), and $r(u)$ and $q(u)$ converge to the monomer values $r=q=1$ 
in the limit of thin cylinders ($u\rightarrow 0$). 

\begin{theorem} \label{laga:theo}
	Let $p\geq 2$ be fixed. 
	Let $r$, $q$ be such that $r^N C_N \rightarrow q$ as in Lemma~\ref{normalization:lem}.
	 The following holds: 
	\begin{enumerate}
		\item There exists a $\gamma_p>0$ such that for $\gamma>\gamma_p$, $r<R_\alpha$. 
		\item The functions $]\gamma_p, \infty[\ni \gamma \mapsto r, q$ are 
			analytic and strictly positive. As $\gamma \rightarrow \infty$, 
			\begin{equation} \label{laga:eq}
				r =1+ O(e^{-\gamma^2}),\quad q= 1+ O(e^{-\gamma^2}).
			\end{equation}
	\end{enumerate}
\end{theorem}

\begin{proof} \emph{1.} Let $0<u\leq v<1$. By Lemma~\ref{actibo:lem}, 
	\begin{align*}
		\alpha_n(u) &= \sum_{m\geq p(n-1)} b_{mn} u^m 
		 \leq  ({u\over v})^{p(n-1)} \alpha_n(v)
	\end{align*}
	with suitable non-negative integers $b_{mn}$. It follows that 
	$u^{p}\thinspace R_\alpha(u)\geq v^{p}\thinspace R_\alpha(v)$. 
	We fix $v$ and let $u\to 0$. Since $R_\alpha(v)\geq r(v)>0$, we obtain that 
	$R_\alpha(u)$ goes to infinity when $u\rightarrow 0$, as expected from Eq.~(\ref{monomervalues}). 
	On the other hand, we know that $C_N\geq \alpha_1 ^N =1$, hence $r(u)\leq 1$. 
	Thus for sufficiently small $u=\exp(-\gamma^2)$, $r(u)\leq 1< R_\alpha(u)$. 
		
   \emph{2.} The positivity of $r$ was proved in Lemma~\ref{normalization:lem}. The positivity of $q$
	is a consequence of $r<R_\alpha$. 
	Now, notice 
	that the power series $A(t,u)$ defines a holomorphic function of two 
	complex variables in the domain $|t|<R_\alpha(|u|)$. 
	For $0 \leq u < \exp(-\gamma_p^2)$, $r(u)$ is the unique solution of
	\begin{equation*}
		A(r(u),u) =1.
	\end{equation*}
	By Lemma~\ref{actibo:lem}, $A(t,0)=t$. 
	Thus the thin cylinder limit corresponds to the point $(r(0),0)=(1,0)$. 
	 We can apply an implicit 
	function theorem for holomorphic functions to obtain the analyticity of $r$. 
	The analyticity of $q$ follows from
	\begin{equation}\label{muu}
		q(u)=\Bigl( \bigl(\partial_t A\bigr)\bigl(r(u),u\bigr) \Bigr)^{-1}.
	\end{equation}
	Both $r(u)$ and $q(u)$ can be extended to holomorphic functions in a complex neighborhood of $u=0$ and 
	take the value $1$ at $0$, whence Eq.~(\ref{laga:eq}).	\qed
%
%
\end{proof}

Theorem~\ref{laga:theo} is the central technical result of the present work, 
as all our results on the one particle density will rely on the condition 
$q >0$.

\subsection{Thermodynamic limits of correlation functions and symmetries}\label{thermo:sec}
In this subsection, we show that Laughlin's state has a unique thermodynamic limit 
(Theorem~\ref{thermolim:theo}) and show that the limiting state 
is periodic in the axial direction, with $p\gamma$ as one of its periods.
 The proof that $p\gamma$ is actually the smallest period is deferred to the next subsection. 
%

The results presented here hold provided $\lim r^n C_n = q>0$, 
i.e., the associated renewal process has finite 
mean. From the previous subsection, we know that this condition is indeed fulfilled on 
sufficiently thin cylinders.

In the following, $\mathcal{A}$ is 
the $C^*$-algebra generated by the fermionic creation and annihilation operators 
$c^*(f), c(g)$, with $f,g\in L^2(\RR\times[0,2\pi/\gamma])$. 
The operators associated with the lowest Landau level basis state $\psi_k$ are 
denoted $c_k^*, c_k$. 

\begin{theorem} [Existence of the thermodynamic limit] \label{thermolim:theo}
	Suppose $r^n C_n \rightarrow q>0$. There is a state $\la \cdot \ra$ such that for 
	every sequence of integers $(a_N)$ such that $a_N \rightarrow \infty$ and $N+a_N \rightarrow \infty$, 
	the states associated with the shifted Laughlin functions $\tilde \Psi_N =t(a_N p\gamma  \ex)^{\otimes 	
	N}\Psi_N$ 
	converge to $\la \cdot \ra$: for all $a\in \mathcal{A}$,
	\begin{equation}\label{cstarlim}
		\la a\ra_N = \frac{1}{C_N} \la \tilde \Psi_N, a \thinspace \tilde \Psi_N \ra 
		\underset {N\rightarrow \infty} {\longrightarrow} \la a \ra.
	\end{equation}
\end{theorem}

\begin{proof}
	For $L=\{\ell_1<..<\ell_r\}\subset \ZZ$, 
	let $c_L:=c_{\ell_1}.. c_{\ell_r}$ and $c_L^*:= (c_L)^*$. It is enough to prove 
	the convergence~(\ref{cstarlim}) for operators $a = c_{L'}^* c_{L}$ with $|L'|=|L|$. 
	 The key idea of the proof is to show a formula similar to the one given in Prop.~\ref{modfunct:prop}
	for the solvable model, and then to use the asymptotics of 
	the normalization constants, just as we did in the proof of Corollary~\ref{thermolim:cor}.
	 Let $b_N:= N+a_N$. 
	We will see that $\la c_{L'}^* c_L\ra _N$ can be written as 
	\begin{equation} \label{repo}
		\la c_{L'}^*c_{L}\ra_N
		=\sum_{n=1}^N \sum_{j=a_N}^{b_N-n} {C_{j-a_N} C_{b_N-j-n}\over C_N} f_{n}(L'-pj,L-pj)
	\end{equation}
	for a suitable $N$-independent family of functions
	$(f_n)_{n\in \NN}$. We use the notation 
	\begin{equation*}
		L-pj = \{\ell_1,..,\ell_r\} - pj = \{\ell_1 - pj, .., \ell_r - pj\}.
	\end{equation*}
	 The $f_n$'s have finite support,
	\begin{equation} \label{support}
		f_{n}(L',L)\neq 0\ \Rightarrow\  L \cup L' \subset\{0,..,pn-p\}, 
	\end{equation}
	and non-negative ``diagonal'' values:
	\begin{equation*}
	 	 \forall L\subset \ZZ:\ f_n(L,L)\geq 0.
	\end{equation*}
	For fixed $j$ and $n$, $C_{j-a_N} C_{b_N-j-n}/C_N\rightarrow qr^n$ 
	due to $C_N r^N \rightarrow q>0$. 
	Thus formally, the right-hand side of Eq.~(\ref{repo}) converges to 
	\begin{equation}\label{repolim}
 		\la c_{L'}^*c_L\ra = \sum_{n=1}^\infty q r^n \sum_{j= -\infty}^\infty f_n(L'-pj ,L - pj),
	\end{equation}
	provided the series converges.

	We prove the theorem in three steps.
	First, we define the auxiliary functions $f_n$ and 
	prove the representation~(\ref{repo}) of the correlation functions. Second,
	we look at ``diagonal'' correlation functions ($L'=L$) and
	 show that the series~(\ref{repolim}) is bounded and equals the limit of correlation functions:
	\begin{equation*}
		\la c_L^* c_L \ra \leq 1, \quad \lim_{N\rightarrow \infty} \la c_L^* c_L \ra _N =
		\la c_L^* c_L \ra. 
	\end{equation*}
	As a last step, we turn to off-diagonal values ($L'\neq L$). We prove that 
	the series~(\ref{repolim}) is absolutely convergent:
	\begin{equation} \label{causs:ineq}
		\sum_{j,n} q r^n |f_n(L'-pj,L-pj)|\leq  \la c_L ^* c_L\ra ^{1/2} 
		 \la c_{L'} ^* c_{L'}\ra ^{1/2}\leq 1
	\end{equation}
	and show that $\la  c_{L'}^* c_L \ra_N\rightarrow  \la  c_{L'}^* c_L \ra$.	
	Note that once we know that Eq.~(\ref{repolim}) defines a state on $\mathcal{A}$, 
	the inequality~(\ref{causs:ineq}) with absolute value bars \emph{outside} the sum  
	is just Cauchy-Schwarz for the state $\la \cdot \ra$. \\

	\emph{1. Representation of correlation functions. }
		 Let $L', L\subset \ZZ$ with $|L'|=|L|$. We start with the representation 
	\begin{equation}\label{sufo}
		\la \Psi_N, c_{L'}^* c_L \Psi_N \ra 
		= \sum_{\vect{m}',\vect{m}}
		 \overline{a_N(\vect{m'})} a_N(\vect{m}) \la \psi_{m'_1}\wedge .. \wedge\psi_{m'_N},
			c_{L'}^* c_L \psi_{m_1}\wedge .. \wedge \psi_{m_N} \ra.
	\end{equation}
	The sum ranges over  $N$-admissible sequences $\vect{m},\vect{m'}$. 
	Suppose $\vect{m}$ and $\vect{m'}$ have common renewal points $s,t$ such that 
	$L\cup L'\subset \{ps,..,pt-p\}$. Then 
	\begin{multline}\label{splitone}
		\la \psi_{m'_1}\wedge .. \wedge\psi_{m'_N},
			c_{L'}^* c_L \psi_{m_1}\wedge .. \wedge \psi_{m_N} \ra \\
		=\prod_{j \in \{1,..,s\}\cup\{t,..,N\}} \delta_{m_j,m'_j} 
		|a_s(\vect{m}_1^s)|^2\thinspace |a_{N-t}(\vect{m}_{t+1}^N-pt)|^2\\
	 	\ \cdot\overline{a_{t-s}(\vect{m'}_{s+1}^t-ps)} a_{t-s}(\vect{m}_{s+1}^t-ps)\\
		\cdot \la \psi_{m'_{s+1}}\wedge.. \wedge\psi_{m'_{t}},c_{L'}^* c_L 
		\psi_{m_{s+1}}\wedge.. \wedge\psi_{m_{t}}\ra.
	\end{multline}
	Let $\mathcal{M}$ be the set of 
	 pairs $(\vect{m},\vect{m'})$ such that 
	\begin{enumerate}\label{enum:calmdef}
	 \item 	$\vect{m}$, $\vect{m'}$ are both $N$-admissible;
	 \item 	$\vect{m}$ and $\vect{m'}$ have no common renewal point $s$ below or above 
		$p^{-1}(L\cup L')$. 
	\end{enumerate}
	By ``$s$ is below $p^{-1}(L\cup L')$'' we mean $L\cup L'\subset \{0,..,ps-p\}$, 
	and we say ''$s$ is above $p^{-1} (L\cup L')$'' when $L\cup L'\subset \{ps,..,pn-p\}$. 
	The set $\mathcal{M}$ consists of the pairs $(\vect{m},\vect{m'})$ for which 
	no simplification of the type~(\ref{splitone}) is possible. 

	$f_N(L',L)$ is defined 
	by the sum~(\ref{sufo}), except that the summation includes only $(\vect{m},\vect{m'})$ 
	from $\mathcal{M}$. 
%
	With this definition, combining (\ref{sufo}) and (\ref{splitone})
	we obtain (\ref{repo}). \\

	\emph{2. ``Diagonal'' correlation functions.} For $L=L'$, the definition of $f_n$ gives 
	$$f_{N}(L,L) = \sideset{}{'} \sum_{\vect{m}} |a_N(\vect{m})|^2 \chi_{L\subset\{m_1,..,m_N\}}$$
	where the sum ranges over  $N$-admissible sequences that 
	are $L$-irre\-duc\-ible. In particular, $f_N(L,L)\geq 0$. Moreover, if $f_N(L,L)\neq 0$, 
	there exists an $N$-admissible sequence $\vect{m}$ such that 
	$L\subset \{m_1,..,m_N\}$, whence $L\subset \{0,..,pN-p\}$. 

	Let $d\in \NN$. From (\ref{repo}) and $f_n(L,L)\geq 0$ we get 
	\begin{equation} \label{eq:step2bound}
		\sum_{n=1}^d  \sum_{j=a_N}^{b_N-j-n} {C_{j-a_N} C_{b_N-j-n}\over C_N} f_n(L-pj,L-pj)
		\leq \la c_L^* c_L \ra_N \leq 1
	\end{equation}
 	If $f_n(L - pj, L - pj )\neq 0 $ 
	we must have $L-pj \subset \{0,..,pn-p\}$, thus only a finite, $N$-independent number of $j$'s contribute 
	to the sum and we can take the limit $N\rightarrow \infty$, which gives 
	\begin{equation*}
		\sum_{n=1}^d \sum_{j=-\infty}^\infty q r^n f_n(L-pj, L-pj) \leq 1.
	\end{equation*}
	Letting $d\rightarrow \infty$, we obtain the bound $\la c_L^* c_L \ra \leq 1$. 
	The proof of $\la c_L^*c_L\ra_N\rightarrow \la c_L^* c_L \ra$ is then 
	completed by an $\epsilon/3$ argument. We leave the details to the reader 
	and mention only a useful inequality on quotients of normalization constants.
	Using the supermultiplicativity of $(C_N)$, 
	$0<C_n r^n \leq 1$ and $C_n r^n \rightarrow q >0$, we get $\inf_n r^nC_n =:c>0$ and 
	\begin{equation*}
		{C_j C_{N-j-n}\over C_N} \leq {C_{N-n}\over C_N} \leq {r^{-(N-n)}\over c r^{-N}} = {1\over c} r^n.
	\end{equation*}
	
	\emph{3. ``Off-diagonal'' correlation functions ($L\neq L'$).} 
	The procedure is similar to step \emph{2.} but the analogue of the 
	bound~(\ref{eq:step2bound}) is slightly 
	more delicate to obtain.
	Let $d\in \NN$. Then
	\begin{align*}
		&\sum_{n=1}^d 	\sum_{j=a_N}^{b_N-n} {C_{j-a_N} C_{b_N-n-j}\over C_N}\, |f_n(L'-pj,L-pj)|\\
		&\qquad \qquad\leq 
			\sum_{n=1}^ N\sum_{j=a_N}^{b_N-n} {C_{j-a_N} C_{b_N-n-j}\over C_N}\, |f_n(L'-pj,L-pj)|\\
		&\qquad \qquad = C_N^{-1}\sum_{\vect{m},\vect{m'}} \bigl|a_N(\vect{m'}) a_N(\vect{m})
		 	\la \psi_{m'_1}\wedge..\wedge\psi_{m_N}, c_{L'}^*c_L \psi_{m_1}\wedge ..\wedge
			 \psi_{m_N}\ra\bigr|\\
		&\qquad \qquad= C_N^{-1}\sum_{K\subset \ZZ,\ |K|=N-|L|} \bigl|a_N(L'\cup K) a_N(L\cup K)\bigr|\\
		&\qquad \qquad\leq C_N^{-1}\bigl( \sum_{ K} |a_N(L'\cup K)|^2 \bigr )^{1/2}
			\bigl( \sum_{K} |a_N(L \cup K)|^2 \bigr )^{1/2}\\
		& \qquad \qquad= \la c_{L'}^* c_{L'} \ra_N ^{1/2}  \la c_{L}^* c_L\ra_N^{1/2}
			\leq 1.
	\end{align*}
	The notation $a_N(L\cup K)$ refers to the amplitude of the 
	increasing sequence obtained by rearranging the elements of $L\cup K$. 
	Letting first $N$ and then $d$  go to infinity, we obtain the bound~(\ref{causs:ineq}).
	The convergence $\la c_{L'}^*c_L\ra _N \rightarrow \la c_{L'}^*c_L\ra$ can be shown with 
	an $\epsilon/3$ argument. \qed
 \end{proof}

\begin{rem} \label{rem:proba} 
 	The representation~(\ref{repolim}) of correlation functions 
	does not lend itself to a simple interpretation. 
	However, it leads to a very nice formula for the one particle density. 
	Let $\hat n_k:=c_k^*c_k$ be the number operator for the lowest Landau level state $\psi_k$. 
	The quantity $f_n(\{k\},\{k\})$ equals $\la u_{X_0}, \hat n_k u_{X_0} \ra$ with the 
	polymer $X_0=\{0,..,pn-p\}$.  Therefore 
	Eq.~(\ref{repolim}) may be rewritten as 
	\begin{equation} \label{eq:densproba}
		\la \hat n_k \ra = \sum_{X} \rho^\mathrm{P}(X) v_X(k)
	\end{equation}
	with $\rho^\mathrm{P}(X) = q \alpha_{N(X)} r^{N(X)}$ and $v_X(k) = \la u_X, 
		\hat n_k u_X\ra / ||u_X||^2$. 
	The sum is over all polymers $X = \{j,..,j+n-1\}$, $j\in \ZZ$, $n\in \NN$. 
	This formula has an intuitive probabilistic 
	interpretation: $v_X(k)$ is the probability of finding a particle in the 
	``site'' $k$, given that $k$ is contained in the polymer $X$ (or, strictly speaking, 
	in $\{p\min X,...,p\max X\}$), which happens with probability 
	$\rho^\mathrm{P}(X)$\footnote{The notation $\rho^\mathrm{P}$ refers to polymer correlation functions as 
	defined in \cite{gruberkunz}. Later, we shall use not only $\rho^\mathrm{P}(X)$ but also 
	$\rho^\mathrm{P}(X,Y)$, see p.~\pageref{page:probacorr}.}.
	
	Together with a similar formula for two-point correlations $\la \hat n_k \hat n_j \ra$, 
	Eq.~(\ref{eq:densproba}) will serve as a useful guide in Sect.~\ref{symclu:sec} when 
	we investigate clustering properties. 
\end{rem}

Now let us turn to the symmetries of Laughlin's state. 
Let $\tau_x$ be the automorphism of the algebra $\mathcal{A}$ 
associated with the magnetic translation $t(\gamma \vx)$. 
Let $\tau_y^a$ be the morphism associated with the  translation $t(a \vy)$ 
in the $y$-direction, and 
$\tau_s$ the morphism induced by the reversal $(s_0 \psi)(z) = \psi(-z)$. 

\begin{proposition}[Symmetries] \label{symmetries:prop}
	The state $\omega(\cdot) = \la \cdot \ra$ of the previous theorem is invariant with respect to 
	reversal, translations in the $x$-direction by multiples of $p\gamma $, and 
	arbitrary translations in the $y$-direction: 
	\begin{equation*}
		\forall n\in \ZZ,\ \forall a \in \RR:\ \omega = \omega \circ \tau_s 
		= \omega\circ \tau_x^{np}
		= \omega \circ \tau_y^a.
	\end{equation*}	
\end{proposition}

\begin{proof} 
	The invariance with respect to $y$-translations is a direct consequence of 
	the fact that $\Psi_N$ has a definite $y$-momentum, see Eq.~(\ref{yinvariance:eq}).
	The reversal invariance follows from the invariance for finitely many particles~(\ref{reversalinv}), 
	see also \cite{swk04}.
	The periodicity with respect to magnetic translations in the direction along the cylinder 
	follows from the representation~(\ref{repolim}) of correlation functions. \qed
\end{proof}

Theorem~\ref{thermolim:theo} and Prop.~\ref{symmetries:prop} lead to 
a simple corollary on the one-particle density:

\begin{corollary}\label{cor:dens}
	Let $\rho_N(z)$ be the one-particle density of Laughlin's state $\Psi_N$. 
	Under the assumptions of Theorem~\ref{thermolim:theo}, the shifted 
	density converges pointwise to the one-particle density $\rho(z)$ of 
	the limiting state $\la \cdot \ra$: 
	\begin{equation} \label{densitysum}
		\lim_{N\rightarrow \infty} \rho_N(z - p\lfloor N/2\rfloor \gamma ) 
		= \rho(z),\quad \rho(z) =\sum_{k=-\infty}^\infty \la \hat n_k \ra |\psi_k(z)|^2.
	\end{equation}
	The density is independent of the coordinate $y=\Im z$ around the cylinder. 
	The density as well as the occupation numbers are 
	periodic and reversal invariant: 
	\begin{alignat*}{2}
		\rho(x+p\gamma) &=\rho(x),&\quad \la \hat n_{k+p} \ra &= \la \hat n_{k} \ra,\\
		\rho(-x) &=\rho(x), & \quad \la \hat n_{-k} \ra & = \la \hat n_{k} \ra.
	\end{alignat*}
\end{corollary}

Note that weak$^*$-convergence of the state $\la \cdot \ra_N$ is replaced with 
pointwise convergence of the one-particle density. This uses the representation of 
the density as a sum of Gaussians with occupation numbers as coefficients 
as in Eq.~(\ref{density:eq}). Due to the good localization of the Gaussians, summation and limits can be interchanged, 
whence Eq.~(\ref{densitysum}).

\subsection{Symmetry breaking and clustering}\label{symclu:sec}

This subsection contains the second part of the main results of this paper: 
Theorem~\ref{symbreak:theo} shows that on sufficiently thin cylinders, 
$p\gamma$ is actually the \emph{smallest} period of 
the limiting state $\la\cdot \ra$ as well as the one-particle density $\rho$ of the 
previous subsection. Thus the state $\la \cdot\ra$ has a larger minimal period than 
the Hamiltonian describing interacting electrons in a magnetic field, 
 whose ground states it is supposed to approximate. 
In this sense, there is symmetry breaking. 

In addition, we prove that the state $\la \cdot \ra$ is mixing with respect 
to magnetic translations in the direction of the cylinder axis (Theorem~\ref{cluster:theo}).

\begin{theorem} [Symmetry breaking]\label{symbreak:theo} 
	Suppose $C_n r^n \rightarrow q>0$. Let $\rho(x)$ be 
	the infinite cylinder density from Cor.~\ref{cor:dens}.
	Then on sufficiently thin cylinders, $p\gamma$ is the smallest 
	period of $\rho(x)$. 
\end{theorem}

\begin{proof}
	Due to Eq.~(\ref{densitysum}) and Lemma~\ref{foutrans:lem}, it is enough to look at the 
	occupation numbers. 
	We will show that 
	\begin{equation} \label{densasy:eq}
		\la \hat n_k \ra = \begin{cases}
					1+ O\bigl(\exp(-\gamma^2)\bigr), & \text{if}\ k \in p\ZZ,\\
					O\bigl(\exp(-\gamma^2)\bigr), & \text{else}.
				\end{cases}
	\end{equation}
	Thus for sufficiently large $\gamma$, the sequence of occupation numbers has $p$ as the 
	smallest period and $p\gamma$ is the smallest period 
	of the one-particle density. 

	The idea behind~(\ref{densasy:eq}) is that the thin cylinder limit is at the 
	same time a monomer limit, see Lemma~\ref{actibo:lem} and p.~\pageref{laga:theo}. 
	The wave function corresponding to a pure monomer system, for $N$ particles, 
	is 
	\begin{equation} \label{taothoufunct}
		u_{\{0\}}\wedge u_{\{1\}}\wedge .. \wedge u_{\{N-1\}} 
			 = \psi_0\wedge \psi_p \wedge.. \wedge \psi_{pN-p}.
	\end{equation}
	In the limit $N\rightarrow \infty$, the corresponding monomer occupation numbers 
	$\la \hat n_k \ra_\mathrm{mon}$ equal $1$ if $k$ is a multiple of $p$, and $0$ otherwise.

	Eq.~(\ref{densasy:eq}) now is a consequence of the following observation: 
	the occupation numbers $\la \hat n_k \ra$ are functions of $v=\exp(-\gamma^2)$ that 
	can be extended to holomorphic functions of $v$ in a complex neighborhood of the 
	monomer point $v=0$. This can be shown with the representation 
	\begin{equation*}
	 	\la \hat n_k \ra = \sum_{n=1}^\infty q r^n \sum_{j=-\infty}^\infty 
		f_n(\{k-pj\},\{k-pj\}) 	=\sum_{n=1}^\infty q r^n g_n(k),
	\end{equation*}
	see Eq.~(\ref{repolim}). $g_n(k)$ is a polynomial of $\exp(-\gamma^2)$, 
	and  $q$ and $r$ are analytic functions of $\exp(-\gamma^2)$. 
	We can adapt the procedure used in the proof of Theorem~\ref{laga:theo} and deduce the
	analyticity of $\la \hat n_k \ra$ for small $u$. \qed
%
%
 \end{proof}

\begin{rems} \emph{1.} The monomer state~(\ref{taothoufunct}) is the Tao-Thouless state,
	corresponding to the reference configuration $\taothou$ on p.~\pageref{taothoum}.
	The fact that Laughlin's wave function for a fixed, \emph{finite} number of particles 
	on very thin cylinders approaches the Tao-Thouless state has been observed by Rezayi and 
	Haldane~\cite{rh94}. The novelty here is twofold: first, the limits $N\rightarrow \infty$ and
	 $\gamma\rightarrow \infty$ can be interchanged; second,  the periodicity survives for small but
	 non-vanishing 	cylinder radius. 

	\emph{2.} If we assimilate orbitals $\psi_k$ with lattice sites $k\in\ZZ$, the 
	restriction 
	of the state $\omega$ to the algebra generated by the number operators $\hat n_k$
	can be described by a probability distribution $P$ on particle configurations on $\ZZ$. 
	Adapting techniques from \cite{am80,agl01}, one can show that 
	the probability measures corresponding to $\omega$ and the shifted states 
	$\omega\circ \tau_x$,..,$\omega\circ \tau_x^{p-1}$ are mutually singular. 
	This result holds provided $r^n C_n \rightarrow q>0$ and the second moment 
	$\sum_n n^2\alpha_n r^n$ is finite. 
	Again, this condition is fulfilled when $\gamma$ is large enough. 
	As a consequence, the $p$ 
	quantum-mechanical states $\omega$, ..,$\omega\circ \tau_x^{p-1}$ 
	are not only \emph{distinct}, but also 
	\emph{orthogonal} in the sense of \cite{bratteli}, Def.~4.1.20. 
\end{rems}

Now we come to clustering properties. Before we state our result in its general form,
let us have a look at two-point correlations $\la \hat n_k \hat n_l\ra$, 
where $\hat n_k = c_k^* c_k $ is the number operator for the state $\psi_k$. 
In the spirit of the remark on p.~\pageref{rem:proba}, $\la \hat n_k \hat n_l\ra$ 
may be interpreted as the probability of finding a particle in the site $k$ and another 
particle in the site $l$. In fact, we have a formula analogous to Eq.~(\ref{eq:densproba}) 
for the one-particle density. Define $\rho^\mathrm{P}(X)$ and $v_X(k)$ as on p.~\pageref{rem:proba}.
Let $v_X(k,l):=\la u_X, \hat n_k \hat n_l u_X \ra/||u_X||^2$ and 
\begin{equation*}\label{page:probacorr}
 	\rho^\mathrm{P}(X,Y) = q r^{N(X)} \alpha_{N(X)} r^{d(X,Y)} C_{d(X,Y)} r^{N(Y)} \alpha_{N(Y)},
\end{equation*}
where $X$ is to the left hand side of $Y$, separated from $Y$ by the distance 
\begin{equation*}
 	d(X,Y):=\min Y-\max X -1 \geq 0.
\end{equation*}
$v_X(k,l)$ is the probability of finding  particles in the sites $k$ and $l$
given that $k$ and $l$ are contained in $[p\min X, p\max X]$, while $\rho^\mathrm{P}(X,Y)$ 
is the probability of finding the rods $X$ and $Y$. 

Suppose that $k<l$. Then the diagonal two-point correlation equals 
\begin{equation} \label{eq:probacorr}
 \la \hat n_k \hat n_l \ra = \sum_{X<Y} \rho^\mathrm{P}(X,Y) v_X(k) v_Y(l) + \sum_X \rho^\mathrm{P}(X) 
			v_X(k.l).
\end{equation}
Again, this formula has an intuitive probabilistic interpretation. The two sums 
correspond to the two different situations that $k$ and $l$ are contained 
in two distinct polymers (first sum) or in the same polymer (second sum). 

When $k$ and $l$ are far apart, the probability that they are in the same 
polymer is small. 
On the other hand, when $X$ and $Y$ are far apart, we may write 
\begin{equation*}
 	\rho^\mathrm{P}(X,Y) \simeq q r^{N(X)} \alpha_{N(X)} qr^{N(Y)} \alpha_{N(Y)} 
		=\rho^\mathrm{P}(X) \rho^\mathrm{P}(Y).
\end{equation*}
Therefore we expect that $\la \hat n_k \hat n_l \ra $ is approximately the same 
as  $\la \hat n_k \ra \la \hat n_l \ra$ when $k$ and $l$ are far apart: 
Laughlin's state inherits clustering properties from the polymer correlations $\rho^\mathrm{P}$.

\begin{theorem} [Clustering] \label{cluster:theo}
	Suppose $r^n C_n \rightarrow q>0$. Then the state 
	$\la \cdot \ra$ of Theorem~\ref{thermolim:theo}
	is mixing with respect to the shifts $\tau_x^{np}, n\in \ZZ$: 
	\begin{equation} \label{clus}
		\forall a,b \in{\cal A}:\  
		\lim_{n\rightarrow \infty}\la a\,\tau_x^{pn}(b)\ra = \la a\ra \la b\ra.
	\end{equation}
\end{theorem}

\begin{proof} We use the notation from the proof of Theorem~\ref{thermolim:theo}.
	It is enough to check~(\ref{clus}) for operators 
	$a=c_{L'}^* c_L$, $b= c_{K'}^* c_K$ with $L,L',K,K'\subset \ZZ$.
	Because of particle number conservation and $y$-invariance,
	the only interesting case is 
	\begin{equation}\label{clunotri:eq}
		|L|=|L'|,\quad |K|=|K'|,\quad \sum_{k\in K'\cup L'} k = \sum_{k\in K\cup L}k .	
	\end{equation}
	In the following we will assume that~(\ref{clunotri:eq}) holds and  
	show that $\la ab\ra -\la a \ra \la b\ra$ 
	is small when $L\cup L'$ is far to the left of $K\cup K'$.
	The main idea is to generalize the formula~(\ref{eq:probacorr})
	for two-point correlations. We will see that 
	\begin{equation}\label{FG:eq}
	   \begin{aligned}
		\la c_{L'}^*c_L c_{K'}^* c_K \ra &= F+G\\
		F & = \sum_{X<Y} q r^{N(X)+ N(Y)} f_X(L',L)\,  C_{d(X,Y)}r^{d(X,Y)}\, f_Y(K',K) \\
 		G &= \sum_X qr^{N(X)} g_X( L',L; K',K)
	    \end{aligned}
	\end{equation}
	where 
	$	d(X,Y)=\min Y - \max X -1$
	and the functions $f_X$, $g_X$ will be defined later. Similarly, 
	\begin{equation} \label{cluprod:eq}
		\la c_{L'}^*c_L \ra \la c_{K'}^* c_K \ra = \sum_{X,Y} q^2 r^{N(X)+ N(Y)} f_X(L',L) f_Y(K',K).
	\end{equation}
	The theorem is proved by making the following arguments precise:
	Suppose $L\cup L'$ and $K\cup K'$ are far away. Then, intuitively, 
	the main contributions to $F$ in~(\ref{FG:eq})	come from polymers 
	$X,Y$ separated by a large distance $d(X,Y)$. Since $r^nC_n\rightarrow q$, we expect that 
	$F$ is close to the righthand side of Eq.~(\ref{cluprod:eq}). 
	The second contribution, $G$, in Eq.~(\ref{FG:eq}) will be bounded by the probability for finding 
	a long polymer, which is small. Thus 
	\begin{equation*}
		\la c_{L'}^*c_L  c_{K'}^* c_K \ra \approx F \approx \la c_{L'}^*c_L \ra \la c_{K'}^* c_K \ra.
	\end{equation*}

	Now we define $f_X$ and $g_X$, and prove Eqs.~(\ref{FG:eq}) and~(\ref{cluprod:eq}).
	The argument resembles step \emph{1.} in the proof of Theorem~\ref{thermolim:theo},
	therefore we shall only give the key elements. 
%
	Let $\mathcal{M'}$ be the set of pairs $(\vect{m},\vect{m'})$ 
	such that:
	\begin{enumerate}
		\item 	$\vect{m}$ and $\vect{m'}$ are $N$-admissible.
	 	\item  $\vect{m}$ and $\vect{m'}$ have no common renewal point 
			$s$ 
		\begin{enumerate}
			\item below or above $p^{-1}(L\cup L'\cup K \cup K')$,
			\item  between $p^{-1}(L\cup L')$ and $p^{-1} (K\cup K')$.
		\end{enumerate}
	\end{enumerate}
	We say that ''$s$ is between $p^{-1}(L\cup L')$ and $p^{-1}(K\cup K')$''  
	if $L\cup L'$ is contained  in $\{0,..,ps-p\}$ and $K\cup K'$ in $\{ps,..,pN-p\}$.  
	The set $\mathcal{M'}$ is a subset of the set $\mathcal{M}$ introduced on p.~\pageref{enum:calmdef}
	for the definition of $f_N$. Let
	\begin{multline*}
	g_N(L',L;K',K)\\:= \sum_{(\vect{m},\vect{m'})\in \mathcal{M'}}
	 \overline{a_N(\vect{m'})} a_N(\vect{m}) 
		\la \psi_{m'_1}\wedge..\wedge \psi_{m'_N}, c_{L'}^* c_L c_{K'}^* c_K 
			\psi_{m_1}\wedge ..\wedge \psi_{m_N} \ra.
	\end{multline*}
	The product rule from Lemma~\ref{prodrule:lem} gives 
	\begin{multline}\label{gclus:eq}
		f_N(L'\cup K',L\cup K)\\
			=\sum_{0< i\leq j< N} f_i(L',L) C_{j-i}f_{N-j}(K'-pj,K-pj)
			+ g_N(L', L;K',K).
	\end{multline}
	For $X=\{j,..,j+N-1\}$, it is convenient to define 
	\begin{align*} 	
		f_X(L', L) &= f_N(L'-pj, L-pj),\\
		 g_X(L',L;K',K) &= g_N(L'-pj, L-pj;K'-pj,K-pj).
	\end{align*}
	Eqs.~(\ref{FG:eq}) and~(\ref{cluprod:eq}) hold with these definitions of $f_X$, $g_X$, 
	as can be seen with Eqs.~(\ref{repolim}) and~(\ref{gclus:eq}).

	Next, we estimate $F-\la c_{L'}^*c_L\ra \la c_{K'}^*c_K\ra$. 
	In view of Eqs.~(\ref{FG:eq}) and~(\ref{cluprod:eq}), this difference is 
	a sum over polymers $X,Y$:
	\begin{equation} \label{Fsum}
		 \sum_{X,Y} q\, r^{N(X)+ N(Y)} f_X(L',L) \Bigl( C_{d(X,Y)}r^{d(X,Y)}-q \Bigr) f_Y(K',K)
	\end{equation}
	with the convention $C_n=0$ for $n<0$. We split the sum in two parts. 
	First, suppose that $X$ is far to the left of $Y$, i.e., $d(X,Y)\geq M$ for some 
	fixed $M\in \NN$. The sum over such $(X,Y)$ is bounded by
	\begin{multline*}
		\sup_{n\geq M} |1-q^{-1} r^n C_n|\, \Bigl(\sum_X qr^{N(X)} |f_X(L',L)|\Bigr)
		\Bigl(\sum_Y q r^{N(Y)} |f_Y(K',K)|\Bigr) \\
		 \leq 	\sup_{n\geq M} |1-q^{-1} r^n C_n|
		\underset{M\rightarrow \infty}{\longrightarrow} 0,
	\end{multline*}
	where we have used Ineq.~(\ref{causs:ineq}). Second, suppose that 
	 $X, Y \subset \ZZ$ satisfy $d(X,Y)\leq M$ and $f_X(L',L)f_Y(K',K)\neq 0$. 
	Then $X$ or $Y$ must be a long polymer, 
	see Fig.~\ref{fig:clust1}. 
	\begin{figure}
		\begin{center}
			\input{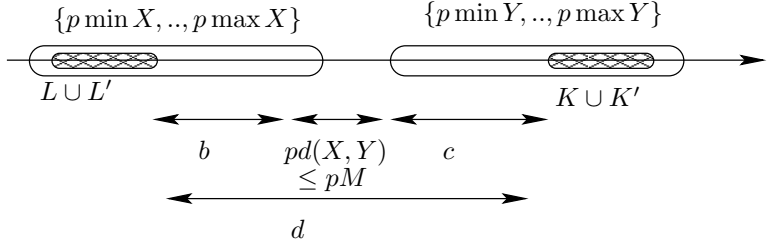}
		\end{center}
		\caption{\small
		Visualization of Ineqs.~(\ref{ineq:cond1}) and~(\ref{ineq:cond2}). \label{fig:clust1}
		If $L\cup L'$ is contained in $[p\min X,p\max X]$ and $K\cup K'$ is contained in 
		$[p\min Y,p\max Y]$, and if the distance $d$ between $L\cup L'$ and $K\cup K'$ is 
		large but the distance $d(X,Y)$ between $X$ and $Y$ is small, $b$ or $c$ must 
		be large.}
	\end{figure}
	More precisely, (\ref{support}) can be used to show that 
	\begin{align}
		p(\max X+ 1) - \max L\cup L' &\geq (d-pM)/2.\label{ineq:cond1}
	\intertext{or}
		\min(K\cup K') - p\min Y &\geq (d-pM)/2 \label{ineq:cond2}
	\end{align}
	must hold. The sum~(\ref{Fsum}) over pairs $X,Y$ such that $X$ satisfies Ineq.~(\ref{ineq:cond1})
	is bounded by 
	\begin{align*}
		\bigl| \sideset{}{'}\sum_X q r^{N(X)} f_X(L',L) \bigr|\, 
			(1+q^{-1})\bigl| \la c_{K'}^* c_K \ra \bigr|
		\leq (1+q^{-1}) \bigl| \sideset{}{'}\sum_X q r^{N(X)} f_X(L',L) \bigr|,
	\end{align*}
	where the prime at the sum refers to the constraint~(\ref{ineq:cond1}).
	By a procedure similar to step \emph{3.} of 
	the proof of Theorem~\ref{thermolim:theo}, one can show that 
	\begin{equation*}
		\bigl| \sideset{}{'}\sum_X q r^{N(X)} f_X(L',L) \bigr| \leq 
		q \sum_{n\geq (d-pM)/2p} nr^n \alpha_n.
	\end{equation*}
	The right-hand side represents the probability 
	that a given point is in a polymer of length greater or equal to 
	$(d-pM)/2p$. 
	The sum~(\ref{Fsum})  over pairs $(X,Y)$ such that $Y$ satisfies Ineq.~(\ref{ineq:cond2}) 
	can be treated in a similar way. 
	In the end, we obtain 
	\begin{equation}\label{ineq:Fab}
		\bigl|F-\la a \ra \la b \ra \bigr|\leq \sup_{n\geq M} |1-q^{-1} r^n C_n|
 		+ 2(1+q) \sum_{n\geq (d-pM)/2p} nr^n \alpha_n.
	\end{equation}
	Choosing $M$ of the order of $d/2p$, we see that $|F-\la a \ra \la b \ra|$ 
	goes to $0$ as $d$ goes to infinity. 
	
	It remains to give a bound on $G$. The idea is to have a closer look at the definition of 
	$g_n$ and to show that if $\vect{m},\vect{m'}$ give a non-vanishing contribution to the 
	sum, at least one of the vectors has no renewal point between $p^{-1}(L\cup L')$ and 
	$p^{-1}(K\cup K')$, see Fig.~\ref{fig:clust2}. This is 
	stronger than the original requirement that $\vect{m}$ and $\vect{m'}$ have no 
	\emph{common} renewal point between $p^{-1}(L\cup L')$ and 
	$p^{-1}(K\cup K')$. As a consequence, one can bound $G$ by the square root of the probability 
	that the space between $p^{-1}(K\cup K')$ and $p^{-1}(L\cup L')$ is covered by one single 
	long polymer.  The square root comes from a Cauchy-Schwarz inequality 
	and accounts for the fact that one of the sequences $\vect{m},\vect{m'}$ 
	may have a renewal point.
	
	\begin{figure}
		\resizebox{8cm}{!}{\input{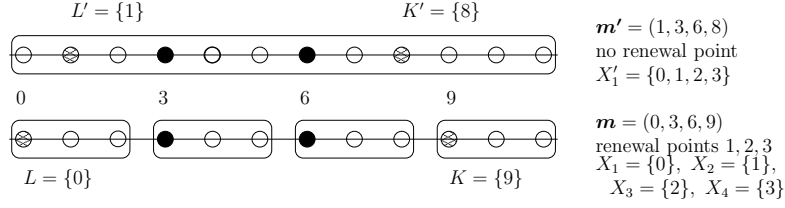}}
		\caption{\small \label{fig:clust2} Evaluation of $G$ in order to estimate 
		$\la c_1^*c_8^* c_0 c_9\ra$ ($p=3$): the sequences $\vect{m'}, \vect{m}$ 
		give a non-vanishing contribution to $g_4(\{1\},\{8\};\{0\},\{9\})$. 
		The vector $\vect{m'}$ has larger partial sums and no renewal points. 
		Each polymer $X$ contributes $N(X)$ particles (filled circles) localized in the discrete 
		volume $\{3\min X,..,3\max X+2\}$ (boxes).
		}
	\end{figure}

	We start by looking at the non-vanishing contributions to $g_n(L',L;K',K)$. 
	Without loss of generality, we may assume $\sum_{k\in L'} k \geq \sum_{k\in L} k$ 
	(otherwise, use $\la a^* \ra = \overline{\la a \ra}$ to interchange $L'$ and $L$).
	Suppose that 
	$\vect{m'},\vect{m}$ are  $N$-admissible  and 
	$$\la c_{L'} c_{K'} \psi_{m'_1}\wedge.. \wedge\psi_{m'_N}, 
		c_L c_K \psi_{m_1}\wedge ..\wedge
			\psi_{m_N}  \ra \neq 0.$$
	We claim that if $s$ is a renewal point of $\vect{m'}$ such that 
	\begin{equation*}
		L\cup L'\subset \{0,..,ps-p\},\quad K\cup K'\subset \{ps,..,pN-p\},
	\end{equation*}
	$s$ is also a renewal point of $\vect{m}$. To see this, let 
	 $M= \{m_1,..,m_N\}$, similarly for $M'$. Then  we must have
	\begin{equation*}
		M'\backslash (K'\cup L') = M\backslash (K\cup L).
	\end{equation*}
	 Intersecting with 
	$\{0,..,ps-p\}$, we obtain
	\begin{equation*}
		(M'\backslash L') \cap \{0,..,ps-p\} = (M\backslash L) \cap \{0,..,ps-p\}.
	\end{equation*}
	In particular,
	\begin{equation*}
	 	\sum_{j=1}^s (m'_j - m_j) = \sum_{k\in L'} k -\sum_{k\in L} k \geq 0.
	\end{equation*}
	 Thus 
	\begin{equation*}
		 0 = \sum_{j=1}^s (m'_j - p(j-1) ) \geq \sum_{j=1}^s (m_j - p(j-1)) \geq 0.
	\end{equation*}
	The inequality on the right-hand side must be an equality, showing that $s$ is a renewal 
	point of $\vect{m}$. 

	Now suppose $\vect{m'},\vect{m}$ give a non-vanishing contribution to the sum defining $g_n$. 
	They are by definition not allowed to have a common renewal point between $p^{-1}(L\cup L')$ and 
	$p^{-1}(K\cup K')$. As we have just shown, this means that $\vect{m'}$ 
	cannot have a renewal point between $p^{-1}(L\cup L') $ and $p^{-1}(K\cup K')$. 
	Thus if $X'_1,..,X'_D$ is the partition of 
	$\{0,..,N-1\}$ 	determined by the renewal points of $\vect{m'}$, there is a rod $X'_j$ 
	such that $[p\min X'_j, p\max X'_j]$ intersects both $L'\cup L$ and $K'\cup K$. 
	It follows that 
	\begin{equation*}
		N(X'_j)\geq \min(K'\cup K)- \max (L\cup L') = d/p.
	\end{equation*}
	In the spirit of step 
	\emph{3.} in the proof of Theorem~\ref{thermolim:theo}, we have 
	\begin{align*}
		&\bigl|\sum_{n=1}^N \sum_{j=0}^{N-n} {C_j C_{N-n-j}\over C_N} g_n(L'-pj,L-pj;K'-pj,K-pj)\bigr|\\
		&\quad \leq C_N^{-1}\sideset{}{'} \sum_{\vect{m},\vect{m'}}
		 \bigl|\overline{a_N(\vect{m})} a_N(\vect{m'}) \la \psi_{m'_1}\wedge.. \wedge\psi_{m'_N}, 
			c_{L'}^*c_L c_{K'}^*c_K 
			 \psi_{m_1}\wedge .. \wedge \psi_{m_N} \ra\bigr| \\
		&\quad  \leq C_N^{-1}\la c_K^*c_K c_L^* c_L\ra_{\Psi_N}^{1/2}\\
		&\quad \qquad \cdot \bigl( 
		\sideset{}{'} \sum_{\vect{m'}} |a_N(\vect{m'})|^2 \la \psi_{m'_1}\wedge .. \wedge \psi_{m'_N},
		c_{L'} ^* c_{L'} c_{K'}^*c_{K'} \psi_{m'_1}\wedge .. \wedge \psi_{m'_N} \ra \bigr )^{1/2},
	\end{align*}
	where $\sum\nolimits'$ is the sum over sequences $\vect{m'}$ having no renewal point between 
	$p^{-1}(L\cup L')$ and $p^{-1}(K'\cup K)$. 
	Taking the limit $N\rightarrow \infty$ after shifting the origin to the middle of the 
	cylinder, we get 
	\begin{equation} \label{ineq:Gab}
		|G| \leq \bigl(q \sum_{n\geq d/p} n r^n \alpha_n\bigr)^{1/2} 
		\underset{d\rightarrow \infty}{ \longrightarrow} 0.\quad  \qed
	\end{equation}
\end{proof}

\noindent \emph{Remark: Rate of convergence.} On sufficiently thin cylinders, we know from 
Theorem~\ref{laga:theo} that $\sum_n \alpha_n t^n$ has a radius of convergence $R_\alpha$ 
strictly larger than $r$. In this case $r^n C_n = q + O\bigl( (r/R_\alpha)^n \bigr)$ and 
Ineqs.~(\ref{ineq:Fab}) and~(\ref{ineq:Gab}) 
show that there is actually \emph{exponential} clustering. 

\begin{acknowledgements}
 E.H.L. thanks the Alexander von Humboldt Foundation for a research award, based at the
Technical University, Berlin. He also thanks the US National Science
Foundation for partial support, grant PHY-0652854. S.J. and R.S. thank the 
DFG for support under grant no. SE 456/7-1.
\end{acknowledgements}

\newcommand{\etalchar}[1]{$^{#1}$}
\providecommand{\bysame}{\leavevmode\hbox to3em{\hrulefill}\thinspace}
\providecommand{\MR}{\relax\ifhmode\unskip\space\fi MR }
\providecommand{\MRhref}[2]{%
  \href{http://www.ams.org/mathscinet-getitem?mr=#1}{#2}
}
\providecommand{\href}[2]{#2}

\end{document}